\documentclass[superscriptaddress,aps,9pt,twocolumn]{revtex4-2}

\usepackage[T1]{fontenc}
\usepackage{newtxtext}
\usepackage{newtxmath}
\usepackage{bbold}

\usepackage{amsmath}
\usepackage{mathtools}
\usepackage{amsfonts}

\usepackage{qcircuit}
\usepackage{braket}

\usepackage{color}
\usepackage{xcolor}
\usepackage[colorlinks,citecolor=blue,linkcolor=blue,urlcolor=blue]{hyperref}

\usepackage{graphicx}
\usepackage[all]{hypcap} 

\usepackage[percent]{overpic}
\usepackage{empheq}
\usepackage[most]{tcolorbox}

\newcommand{\Tr}[1]{\text{Tr} #1}

\begin{document}

\title{Dynamic Cooling on Contemporary Quantum Computers}

\author{Lindsay Bassman Oftelie}
\affiliation{NEST, Istituto Nanoscienze-CNR and Scuola Normale Superiore, I-56127 Pisa, Italy}

\author{Antonella De Pasquale}
\affiliation{Istituto Italiano di Tecnologia, Graphene Labs, Via Morego 30, I-16163 Genova, Italy}

\author{Michele Campisi}
\affiliation{NEST, Istituto Nanoscienze-CNR and Scuola Normale Superiore, I-56127 Pisa, Italy}

\begin{abstract}
We study the problem of dynamic cooling whereby a target qubit is cooled at the expense of heating up $N-1$ further identical qubits, by means of a global unitary operation. A standard back-of-the-envelope high temperature estimate establishes that the target qubit temperature can only be dynamically cooled by at most a factor of $1/\sqrt{N}$. Here, we provide the exact expression for the minimum temperature to which the target qubit can be cooled and reveal that there is a crossover from the high initial temperature regime where the scaling is in fact $1/\sqrt{N}$ to a low initial temperature regime where a much faster scaling of $1/N$ occurs. This slow $1/\sqrt{N}$ scaling, which was relevant for early high-temperature NMR quantum computers, is the reason dynamic cooling was dismissed as ineffectual around 20 years ago; the fact that current low-temperature quantum computers fall in the fast $1/N$ scaling regime, reinstates the appeal of dynamic cooling today. We further show that the associated work cost of cooling is exponentially more advantageous in the low temperature regime. We discuss the implementation of dynamic cooling in terms of quantum circuits and examine the effects of hardware noise.  We successfully demonstrate dynamic cooling in a 3-qubit system on a real quantum processor. Since the circuit size grows quickly with $N$, scaling dynamic cooling to larger systems on noisy devices poses a challenge.  We therefore propose a suboptimal cooling algorithm, whereby relinquishing a small amount of cooling capability results in a drastically reduced circuit complexity, greatly facilitating the implementation of dynamic cooling on near-future quantum computers.  
\end{abstract}

\maketitle
\section{Introduction}\label{sec:intro}
Quantum computers offer massive advantages over classical computers in terms of execution time and memory efficiency for a subset of problems, such as optimization and simulation \cite{preskill1998reliable, bassman2021simulating}.  While various physical implementations of quantum computers are still being explored (e.g., superconducting circuits, ion traps, neutral atoms), all must fulfill a fundamental set of requirements \cite{divincenzo2000physical}.  One of these requirements is the ability to initialize the quantum bits, or qubits, into a pure, fiducial quantum state.
Furthermore, pure ancilla qubits will be required for fault-tolerant quantum computers to perform quantum error correction \cite{knill1998resilient, aharonov1997fault,preskill1998reliable}.  The preparation of pure state qubits is therefore a key hurdle in the successful implementation of quantum computers now and in the future.

\begin{figure}[b]\label{fig:schematic}
\centering
\begin{overpic}[width=0.98\columnwidth]{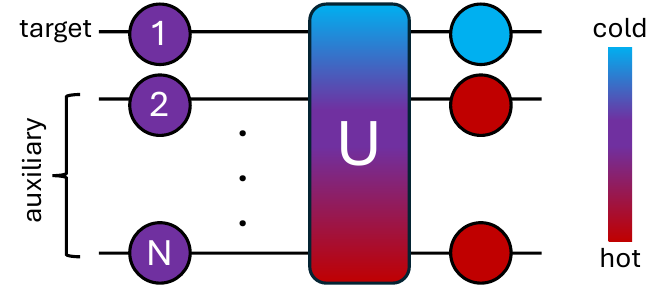}
\end{overpic}
\caption{Schematic of the dynamic cooling of $N$ identical qubits.  Here, the target qubit is cooled at the expense of heating up the auxiliary qubits via the application of a global unitary operator $U$.}
\end{figure}

The problem of initializing a large set of qubits into a pure state was first studied in the context of generating highly polarized qubits in nuclear magnetic resonance (NMR) systems to improve signal-to-noise ratios \cite{morris1979enhancement, sorensen1989polarization, schulman1998scalable}.  Since large polarization in qubits can be obtained by cooling them down to very low temperatures, scientists began to explore techniques to cool qubits below temperatures that can be achieved with direct, physical cooling methods (e.g., cooling with lasers or large magnetic fields).  Schulman and Varizani were the first to propose effective cooling of qubits for quantum computation via the application of certain logic gates on the qubits \cite{schulman1999molecular}, which following Ref. \cite{allahverdyan2011thermodynamic}, we refer to as \textit{dynamic cooling}. Their proposal, based on entropy manipulation in a closed system, cools a subset of qubits (e.g., a single target qubit) at the expense of heating the others by performing unitary operations on the entire set of qubits, see Figure \ref{fig:schematic}.

In the high initial temperature regime, which was relevant for the NMR-based quantum computers available at the time dynamic cooling was proposed, the Shannon bound establishes that the target qubit can be dynamically cooled by a factor of at most $1/\sqrt{N}$, where $N$ is the total number of qubits \cite{boykin2002algorithmic, fernandez2004algorithmic}. 
This slow scaling led to the dismissal of dynamic cooling as an impractical method for cooling qubits, and gave thrust to further research aimed at beating Shannon's bound. Since the bound holds for closed systems, subsequent proposals extended the scenario to open systems by allowing a subset of qubits to interact with the environment (i.e., a heat bath), thereby achieving cooling beyond Shannon's bound \cite{boykin2002algorithmic, fernandez2004algorithmic}. Such techniques are usually referred to as heat bath algorithmic cooling (sometimes simply algorithmic cooling) \cite{schulman2005physical, fernandez2005paramagnetic,  baugh2005experimental, rempp2007cyclic, kaye2007cooling, ryan2008spin, elias2011semioptimal, ticozzi2014quantum, brassard2014prospects, brassard2014experimental, raeisi2015asymptotic, rodriguez2016achievable, atia2016algorithmic, rodriguez2017correlation, raeisi2019novel}.

However, quantum computing technology has undergone a dramatic revolution in the last two decades, with high-temperature NMR quantum computers falling out of favor 
as newer models that operate at very low temperatures (e.g., superconducting circuits, ion traps) have shown great promise \cite{arute2019quantum, bruzewicz2019trapped}.  In tandem, thanks to the development of quantum thermodynamics, much interest has grown within the scientific community in regard to the possible advantage, in terms of energy consumption, of quantum technology in general and quantum computing in particular \cite{Auffeves22PRXQ3}.  

Here, we re-examine dynamic cooling in light of the scientific and technological advances that have been achieved since its inception over two decades ago. We consider a set of $N$ identical qubits, each initially in thermal equilibrium at some initial temperature $T$, that undergoes dynamic cooling via a global unitary transformation $U$, schematically represented in Figure \ref{fig:schematic}. After such a transformation, the ground and excited state populations of the target qubit change, thereby affecting a change in its temperature. We analytically solve the problem of finding the minimum final temperature $T'$ that can be achieved as a function of initial temperature $T$, qubit resonant frequency $\omega$, and total number of qubits $N$. This allows us to unveil a crossover from the expected $1/\sqrt{N}$ scaling at high $T$ to a much faster, unexpected, $1/N$ scaling at low $T$.

We also provide an analytical expression for the minimal work cost associated to maximal cooling and show that it scales linearly with $N$ (i.e., it is extensive) and displays distinct behaviours at low and high temperature. While it vanishes like  $1/T$ in the high $T$ regime, it vanishes exponentially as $e^{-1/T}$ in the low $T$ regime.  These results evidence that dynamic cooling behaves very differently at high and low initial temperatures. In particular, at low $T$ it is much more effective in terms of system-size scaling and energy cost. 

Since current quantum computers operate in the low $T$ regime (unlike early NMR quantum computers), these results reinstate the appeal of dynamic cooling for generating pure state qubits for quantum computation. Given this renewed viability, we discuss the the implementation of dynamic cooling in terms of quantum circuits, and examine the effect of noise on cooling on near-term quantum computers.  We successfully demonstrate dynamic cooling on a real quantum processor on a system of $N=3$ qubits.  While scaling dynamic cooling up to larger systems on noisy quantum computers is a challenge due to the rapid growth of circuit size with $N$, we demonstrate how this can be overcome by accepting a suboptimal cooling scheme, whereby increased cooling can be achieved at a fixed (low) circuit complexity as the system size is increased.  Our re-examination of dynamic cooling suggest that it is a promising technique for preparing pure state qubits on near-future quantum computers.

\section{Maximal Cooling}\label{sec:maximalcooling}
The initial state of the global system reads
\begin{align}
\rho= \bigotimes_{i=1}^N \left(\frac{e^{-\beta H_i}}{Z(\beta)}\right)
\end{align}
where $\beta= 1/(k_B T)$, $k_B$ is Boltzmann's constant and $Z(\beta)=\Tr\, e^{-\beta H_i}= 2\cosh(\beta \omega /2)$ is the partition function of any of the qubits, whose Hamiltonian $H_i=\hbar \omega \sigma_z^i/2$, is here expressed in terms of the Pauli operator $\sigma_z^i$, the reduced Plank's constant $\hbar$ and the resonant frequency $\omega$. All qubits are assumed to have the same resonant frequency.

Let $\ket{i_1 \hspace{1 mm} i_2 \dots  i_N}$ denote the tensor product $\ket{i_1}_1 \otimes  \ket{i_2}_2 \otimes \dots \otimes \ket{i_N}_N$ of the eigenvectors $\ket{i_k}_k$, of the operators $\sigma_z^{k}$, $k=1,2 \dots N$, where $i_k=0$ ($i_k=1$) denotes the ground (excited) state of qubit $k$. We assign qubit $k=1$ to be the target qubit to be cooled and let $P_0 = e^{\beta \hbar \omega/2}/Z(\beta)$ and $P_1=e^{-\beta \hbar \omega/2}/Z(\beta)=1-P_0$ denote the initial ground and excited state populations of the target qubit, respectively.  Similarly, let $P'_0,P'_1=1-P'_0$ denote the final ground and excited state populations of the target qubit after application of the cooling unitary.

To maximally cool the target qubit, the goal is to minimize $P'_1$ over all possible global unitaries $U$.  This problem is equivalent to finding the set of unitaries that minimizes the expectation value of the final energy of the target qubit $u'=\hbar\omega (P'_1-P'_0)=\hbar\omega (2P'_1 - 1)$. Thus, we must solve the minimization problem:
\begin{align}
\label{eq:minimization_equation}
u'&= \min_U \Tr K U \rho U^\dagger
,\quad
K =  H_1 \otimes \mathbf{1} \otimes \dots \otimes \mathbf{1}
\end{align}
where $K$ is the target qubit Hamiltonian expressed in the Hilbert space of the total system.  This problem is formally identical to finding the ergotropy of a driven system (ergotropy is the maximum extractable work from a quantum system) \cite{allahverdyan2004maximal}. The only difference is that solving for the ergotropy addresses the total system energy, setting $K$ in Eq. \ref{eq:minimization_equation} to the total system Hamiltonian. Here, we address the energy of a subsystem (i.e., the target qubit), using its Hamiltonian for $K$.  Since the specific form of the Hamiltonian is irrelevant to the objective minimization problem, we may borrow techniques used to compute the ergotropy and directly apply them to our problem. 

To do so, we note the critical fact that for our system, the initial state $\rho$ commutes with the Hamiltonian $K$.  As discussed in Ref. \cite{allahverdyan2004maximal}, in such a case the optimization is particularly simple:  the minimum is achieved when $U$ is a permutation matrix that maps the eigenstate of $\rho$ with largest eigenvalue to the eigenstate of $K$ with smallest eigenvalue, the eigenstate of $\rho$ with second largest eigenvalue to the eigenstate of $K$ with second smallest eigenvalue, etc. 
In other words, if $e_i$ are the eigenenergies of $K$, then if we order the eigenvalues $p_{i}$ of $\rho$ in non-increasing fashion, that is
\begin{align}
  p_i \ge p_k \quad \text{for} \quad {i<k}\, ,
  \label{eq:p-order}
\end{align}
the minimizing unitary $U$ in Eq. \ref{eq:minimization_equation} is one that performs the permutation $\sigma$ such that 
\begin{align}
  e_{\sigma(i)} \le e_{\sigma(k)} \quad \text{for} \quad {i<k}~.
  \label{eq:e-order}
\end{align} 
Note that this holds even in the case of degenerate spectra.  Indeed, in our case the spectrum of $K$ is highly degenerate, with only two distinct eigenvalues: $e_i = \pm\hbar\omega/2$. States of the form $\ket{0 \hspace{1 mm} i_2 ... i_{N}}$, which are half of the total $2^N$ states, have an eigenenergy of $K$ equal to $-\hbar \omega/2$, while states of the form $\ket{1 \hspace{1 mm} i_2 ... i_{N}}$ have an eigenenergy of $K$ equal to $+\hbar \omega/2$.  Maximal cooling can thus be implemented by mapping the half of states with the highest occupation probabilities to the half of states with the lower eigenenergy $e_i = -\hbar \omega/2$. Due to the large degeneracy in the spectrum of $K$ (as well as degeneracy in the spectrum of $\rho$), there will be many distinct permutations $\sigma$, and hence many distinct unitaries $U$, that achieve maximal cooling.  Illustrative examples for $N=3,4$ are provided in Appendix \ref{sec:illustrative_exs}.

The maximum amount by which the target qubit can be cooled is determined by calculating its final excited state population $P'_1$.  Note that $P'_1$ is simply the sum of the final occupation probabilities of the states $\{\ket{1 \hspace{1 mm} i_2 ... i_{N}}\}$, which are the exact set of states to which the lowest half of probabilities are mapped. Therefore, to compute $P'_1$ we simply generate a list of all the occupation probabilities in non-decreasing order and sum the first half of the list. 

To do so, note that the occupation probability of a state with $k$ bits set to $0$ is given by $(1-P_1)^kP_1^{N-k}$ (where $P_1$ is given by the initial temperature of the qubits), and there will be $\binom{N}{k}$ states with this probability.  States with more bits set to $0$ (higher $k$) have higher initial occupation probabilities.  Therefore, we can generate a list of the probabilities in non-decreasing order by appending the $\binom{N}{k}$ probabilities of value $(1-P_1)^kP_1^{N-k}$ to the list as we increase $k$ from 0 to $N$. Summing the first half of this list will give $P'_1(P_1,N)$.

When $N$ is odd, there are an even number of distinct values of $k$ ranging from $k=0$ to $k=N$.  The number of probabilities for the first half of $k$ values ($k=0, ..., \lfloor N/2 \rfloor$) is equal to the number of probabilities for the second half of $k$ values ($k=\lceil N/2 \rceil, ..., N$).  Thus,
\begin{align}
P'_1(P_1,N) &= \sum_{0 \leq k<N/2}  {N\choose{k}}(1-P_1)^k P_1^{N-k}, \quad \text{for odd } N. 
\label{eq:y_Nodd}
\end{align}

The calculation is slightly more complicated when $N$ is even, since now, dividing the list of probabilities in half involves splitting in half the degenerate group of probabilities where $k=N/2$.  This means that we must add to Eq. \ref{eq:y_Nodd} half of the $\binom{N}{N/2}$ degenerate probabilities with value $(1-P_1)^{N/2}P_1^{N/2}$:

\begin{align} \label{eq:y_Neven}
P'_1(P_1,N)	&= \sum_{0 \leq k<N/2}  {N\choose{k}}(1-P_1)^kP_1^{N-k} \nonumber\\
            &+\frac{1}{2}{{N}\choose{N/2}} (1-P_1)^{N/2}P_1^{N/2},  \quad \text{for even } N. 
\end{align}

An intriguing observation is that if we start from an odd number of qubits, adding one more qubit will not increase maximal cooling:
\begin{align}
P'_1(P_1,2s-1) = P'_1(P_1,2s) \qquad s \in \mathbb{N}
\label{Eq.even=odd}
\end{align}
(proof provided in Appendix \ref{sec:proof}).  This generalizes the fact that a total of at least three (identical) qubits is required to obtain some cooling \cite{solfanelli2022quantum}. To see this, note that with a total of one qubit no cooling is possible by means of a unitary manipulation, so Eq. \ref{Eq.even=odd} implies that cooling with a total of two identical qubits is likewise impossible; a minimum of three qubits is required for dynamic cooling.

\begin{figure}\label{fig:P1prime_vs_P1}
\centering
\begin{overpic}[width=\columnwidth]{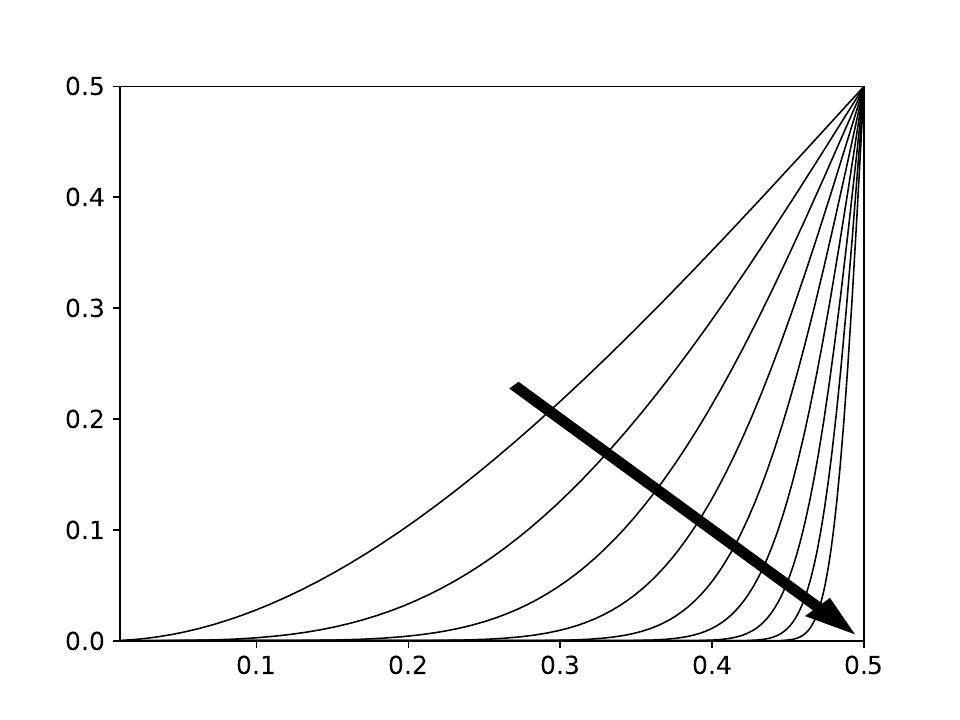}
\put(0,35){\rotatebox{90}{$P'_1$}}
\put(48,0){$P_1$}
\put(51,36){{\textbf{$N \uparrow$}}}
\end{overpic}
\caption{Final probability of the excited state of the target qubit after maximal cooling $P'_1$ versus its initial probability $P_1$, for increasing numbers of total system qubits $N=2^2, 2^3, ..., 2^{10}$.}
\end{figure}

Figure~\ref{fig:P1prime_vs_P1} shows $P'_1$ as a function of $P_1$ for increasing system sizes $N$.  Note that $P'_1(P_1,N)$ is an increasing function of $P_1$, meaning that the larger the initial temperature, the larger the final temperature, which agrees with intuition. Note also that, in the interval $[0,1/2[$, $P'_1(P_1,N)$ is a decreasing function of $N$, namely, the larger $N$ the higher the cooling, in agreement with what one would expect. We have
\begin{align}
\lim_{N\to \infty} P'_1(P_1,N) = 0, \qquad P_1 \in [0,1/2[
\end{align}
meaning that as long as the initial temperature is finite and non-negative, by increasing $N$ one can cool the target qubit arbitrarily close to zero temperature. Note however the crucial fact that $P'_1(1/2,N)=1/2$ for any $N$. This is because any unitary evolution leaves the completely mixed state unaltered; no cooling is possible if the initial temperature is infinite, regardless of $N$. This  constraint is responsible for the low $1/\sqrt{N}$ scaling at high temperature, which will be discussed below.

Using the relation between the initial excited state population $P_1$ and temperature $T$
\begin{align}\label{eq:p_to_T}
\frac{k_B T}{\hbar \omega} = \frac{1}{\ln (1/P_1-1)},
\end{align}
as well as the analogous relation between the final, minimal excited state population $P'_1$ and temperature $T'$, we can write the final minimal temperature as a function of the initial temperature $T$ as:
\begin{align}\label{eq:Tf_vs_Ti}
T' &= \frac{\hbar \omega}{k_B} \frac{1}{\ln( [P'_1(\frac{1}{e^{\hbar \omega/(k_B T)}+1},N)]^{-1}-1)}~.
\end{align}

Here the expression \textit{final temperature} is not being used in a strictly thermodynamic sense, i.e., to denote the temperature of the thermal bath surrounding the qubit, but rather it is being used in an "effective'' sense, i.e., to the denote the temperature that the bath would have if it were in equilibrium with the qubit.

\begin{figure}\label{fig:Tf_vs_Ti}
\centering
\begin{overpic}[width=0.94\columnwidth]{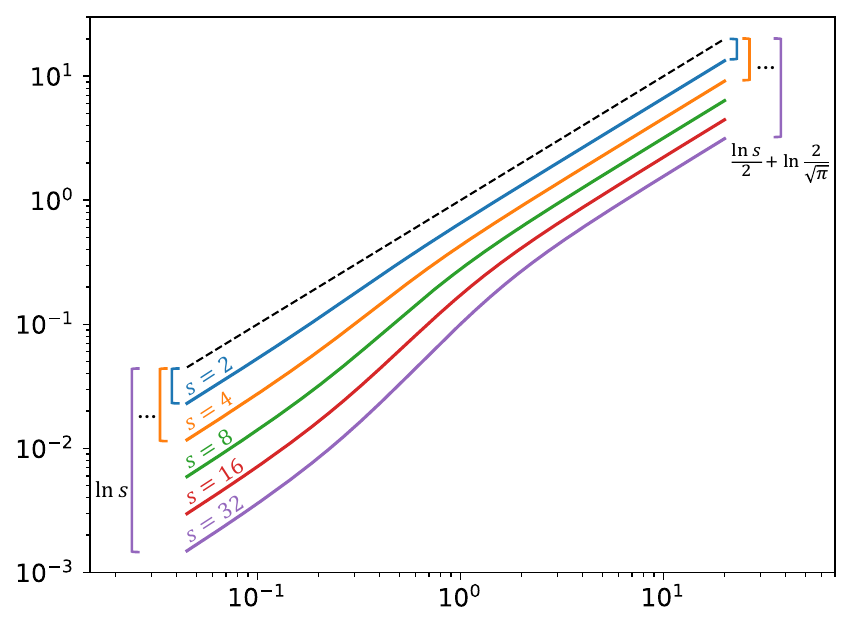}
\put(-3,27){\rotatebox{90}{$k_BT'/\hbar\omega$}}
\put(45,-3){$k_BT/\hbar\omega$}
\end{overpic}
\caption{Log-log plot of the minimum final temperature $T'$ versus the initial temperature $T$ for various system sizes $s=\frac{N}{2}$.  The black dashed line plots $T' = T$ (slope equal to $1$) to guide the eye in seeing the amount of cooling that occurs.   In the low-$T$ regime (bottom-left), curves for various system sizes are parallel with a slope of $1$, implying a linear relationship between $T'$ and $T$; and the curve for each system size $s$ has a vertical shift of $\ln s$ from the black dashed line. In the high-$T$ regime (top-right), curves for various system sizes are also parallel with a slope of $1$, but now the curve for each system size $s$ has a vertical shift of $\frac{\ln s}{2} + \ln(\frac{2}{\sqrt{\pi}})$ from the black dashed line.}
\end{figure}

Figure \ref{fig:Tf_vs_Ti} shows a log-log plot of $T'$ versus $T$ for various system sizes $N=2s$.  The black dashed line plots $T' = T$ to guide the eye in seeing the amount of cooling that occurs.  
In both the low-$T$ regime and the high-$T$ regime, there is a linear relationship between $T'$ and $T$ (the slope of the log-log plots is 1) but the coefficient of proportionality (i.e. the vertical shift of the plots) scales differently with $s$ in the two regimes.

In the high-$T$ regime, $P_1$ is close to $1/2$, hence we Taylor expand $P'_1(P_1,2s-1)=P'_1(P_1,2s)$ around $P_1=1/2$ to obtain
\begin{align}
P'_1 &=  1/2 + c_s (P_1-1/2) + O[(P_1-1/2)^2] \\
c_{s}&= 2^{2-2s} \sum_{k=0}^{s-1} {{2s-1}\choose{k}}(2s-2k+1).
\end{align}
Expanding the expression $e^{\beta \hbar \omega}=(1-P_1)/P_1$ to first order around $P_1 = 1/2$, we obtain $\beta \hbar \omega \simeq 4(1/2-P_1) $. Similarly, for the inverse final temperature, we have $\beta' \hbar \omega \simeq 4(1/2-P'_1) \simeq   c_s 4(1/2-P_1) \simeq  c_s \beta \hbar \omega$, or $T' \simeq T/c_s$. This explains the linear relationship between  $T'$ and $ T$. 
It can be proven that $c_s= 2^{2-2s} s {{2s-1}\choose{s}}$, then, using Stirling's approximation, $N! \simeq \sqrt{2\pi N} (N/e)^N$,  one finds that $c_{s}\simeq (2/\sqrt{\pi}) \sqrt{s}$ in the large $s$ limit (see Appendix \ref{sec:cs_deriv}). 
Therefore:
\begin{align}
T' \simeq  \frac{\sqrt{\pi}}{2} \frac{T}{\sqrt{s}}
= \sqrt{\frac{\pi}{2}}   \frac{T}{\sqrt{N}}  \qquad  \text{for } \quad k_B T  \gg \hbar \omega~.
\label{eq:Tf-highT}
\end{align}
Note that  $T'> T/\sqrt{N}$  because $\sqrt{\pi/2}>1$, which means that Shannon's bound is obeyed as expected, but not saturated. Finding that the scaling $1/\sqrt{N}$ is realized in the high T regime (as opposed to just a theoretical bounding limit) is \textit{per se} a non trivial result. This slow scaling is clearly visible in the top right corner of Figure \ref{fig:Tf_vs_Ti}.

In the low-$T$ regime, we have $P_1\ll 1$. Taylor expansion of $P'_1(P_1,2s-1)=P'_1(P_1,2s)$ around $P_1=0$ gives
\begin{align}
    P'_1 &= a_s P_1^s + O(P_1^{s+1})
   ,\quad  a_s = {{2s-1}\choose{s}}
\end{align}

For small $P_1$, we have $e^{-\beta \hbar \omega}=P_1/(1-P_1)\simeq P_1 $, and a small $P'_1$. Therefore, $e^{-{\beta}' \hbar \omega}=P'_1 /(1-P'_1)\simeq P'_1 \simeq a_s P_1^s = e^{-s \beta \hbar \omega+ \ln a_s}$. 
This implies $\hbar \omega \beta' \simeq s \hbar \omega \beta - \ln a_s$.
Using Stirling approximation we obtain 
$ a_s \simeq s\ln 4 + O(\ln s)$ (see Appendix \ref{sec:as_deriv}), hence $\hbar \omega \beta' \simeq s ( \hbar \omega \beta- \ln 4)$.
At low temperature (i.e., $\hbar \omega \beta \gg1 $)  the term $\ln 4$ is negligible, therefore $\beta' \simeq s \beta$, or:
\begin{align}
T' \simeq  \frac{T}{s} = 2 \frac{T}{N} \qquad  \text{for } \quad k_B T  \ll \hbar \omega.
\label{eq:Tf-lowT} 
\end{align}
This reveals that in the low-$T$ regime, a much faster $1/N$ scaling holds for dynamic cooling. This superior scaling is clearly visible in the bottom left corner of Figure \ref{fig:Tf_vs_Ti}.

A characteristic value of $k_B T/\hbar \omega$ for contemporary quantum computers based on superconducting qubits, ion traps, or neutral atoms is $\simeq 0.2$, which places them within the start of the low $T$ regime. For example, current superconducting qubit quantum computers typically operate at $\omega = 5$ GHz and $P_1 = 0.01$, which equates to an initial temperature of $T = 8.3$ mK.  Given these values, we find $T' = 2.1$ mK for $s=5$ ($N=9,10$), which is slightly above the scaling value of $T/s = 8.3/5$ mK $= 1.66$ mK. However, in accordance with our analysis above, the estimate $T/s$ becomes better and better as $N$ increases and/or as $T$ decreases further.  

\section{Minimal Work}\label{sec:minimalwork}
Due to the large degeneracy of the spectrum of $K$ (defined in Eq. \ref{eq:minimization_equation}), there is a great number of distinct permutations that achieve the desired ordering of eigenvectors for maximal cooling. A natural question is then, which among all these permutations have the smallest cost in terms of energy injection into the system, i.e., the work performed on the  system, given by
\begin{align}
\label{eq:work}
W = \Tr \left[H (U  \rho U^\dagger- \rho ) \right] 
\end{align}
where $H = \bigotimes _iH_i$ denotes the total system Hamiltonian. We recall that, since the initial state is passive, we have $W\geq 0$. When $U$ realizes a permutation $\sigma$, Eq. \ref{eq:work} boils down to
\begin{align}
W =  \sum_i E_i  (p_{\sigma(i)}- p_i ) 
\label{eq:work-permutation}
\end{align}
where $E_i$ are the eigenvalues of $H$. Minimal work cost is thus determined by the following minimization problem:
\begin{align}
    \overline{W} = \min_{\sigma \in \mathcal{C}} \sum_i E_i  (p_{\sigma(i)}- p_i ) 
    \label{eq:min-W}
\end{align}
where $\mathcal{C}$ denotes the set of permutations that realize maximal cooling.

Solving this further minimization problem is straightforward. As described above, in order to achieve maximal cooling it is sufficient to map the half of states with the highest occupation probabilities to the set of states $\{\ket{0 \hspace{1 mm} i_2 ... i_{N}}\}$.  To simultaneously achieve minimal work cost, within this set of states the highest probability should be assigned to the state with the lowest \textit{total system} energy, the second highest probability should be assigned to the state with the second lowest total system energy, etc. The probabilities should be mapped in an analogous way for the other half of states in the set $\{\ket{1 \hspace{1 mm} i_2 ... i_{N}}\}$. This works because states with lower total system energies have higher \textit{initial} occupation probabilities by definition.  So assigning the highest \textit{final} probability to the state with the lowest total system energy within each half-list minimizes the differences of the initial and final probabilities $p_{\sigma(i)} - p_i$ in Eq. \ref{eq:work-permutation}, thereby minimizing work (see Appendix \ref{sec:illustrative_exs} for more details).

\begin{figure}\label{fig:rescaled_W_vs_P1}
\centering
\begin{overpic}[width=0.94\columnwidth]{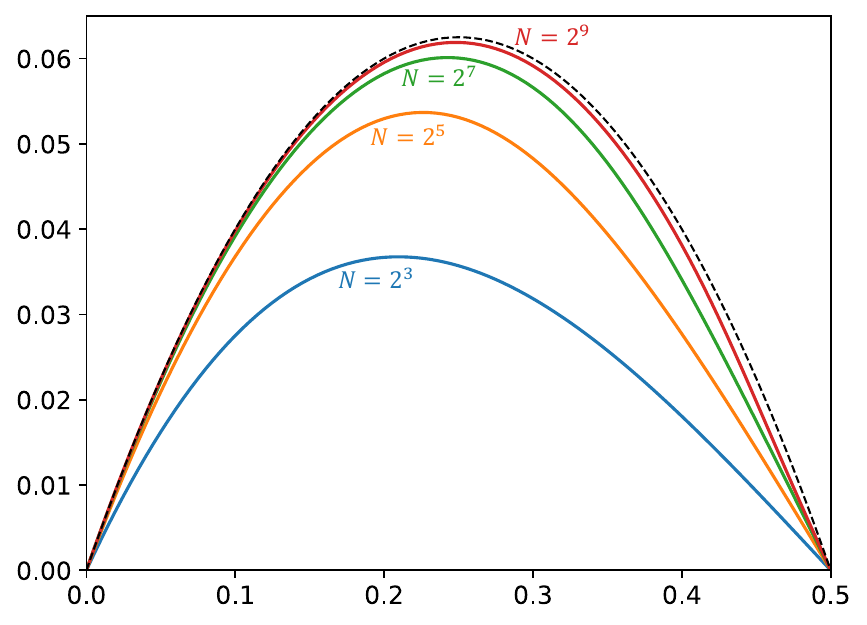}
\put(-4,36){\rotatebox{90}{$\overline{W}/N$}}
\put(50,-2){$P_1$}
\end{overpic}
\caption{Rescaled minimum work versus $P_1$ for various system sizes $N$. Dashed black line is Eq. \ref{eq:rescaled_min_W}.}
\end{figure}

Computing the minimal value of work $\overline{W}$ that must be invested to obtain maximal cooling is conceptually a simple task, but, in practice, it presents some challenges. Note that due to memory limitations, writing the $2^N$ dimensional arrays that list the energy eigenvalues $E_i$, and the populations $p_i$, $p_{\sigma(i)}$ quickly becomes intractable as $N$ increases (on a desktop computer this already happens around $N\simeq 26$). We overcome this bottleneck by exploiting the sparsity of these arrays which allows us to encode the relevant information into arrays whose sizes scale linearly, thereby allowing the evaluation of $\overline{W}$ for $N$ into the thousands.

Figure \ref{fig:rescaled_W_vs_P1} shows the rescaled work $\overline{W}/N$ as function of $P_1$ for various $N$. For $P_1 \in [0,1/2]$, our numerical calculations cleary evidence that as $N$ is increased the solid curves approach the black dashed curve, which plots the following analytic expression:
\begin{align}
\label{eq:rescaled_min_W}
\lim_{N\to \infty } \frac{\overline{W}(P_1,N)}{N} \doteq \bar{w}(P_1) 
= \frac{\hbar \omega}{2} (P_1-2P_1^2). 
\end{align}
The minimal work is extensive which evidences a trade-off between cooling power and energetic cost: the further one cools a qubit, the more energy one must expend. At low $T$, this trade-off is balanced, as the product $\overline{W} T' \sim 2\bar{w} T$ is of order 1.  However, in the high $T$ limit, the trade-off is disadvantageous because the product $\overline{W} T' \sim (\sqrt{\pi}/2) \bar{w} T \sqrt{N} $ scales like $\sqrt{N}$.

\begin{figure}\label{fig:rescaled_W_vs_T}
\centering
\begin{overpic}[width=0.94\columnwidth]{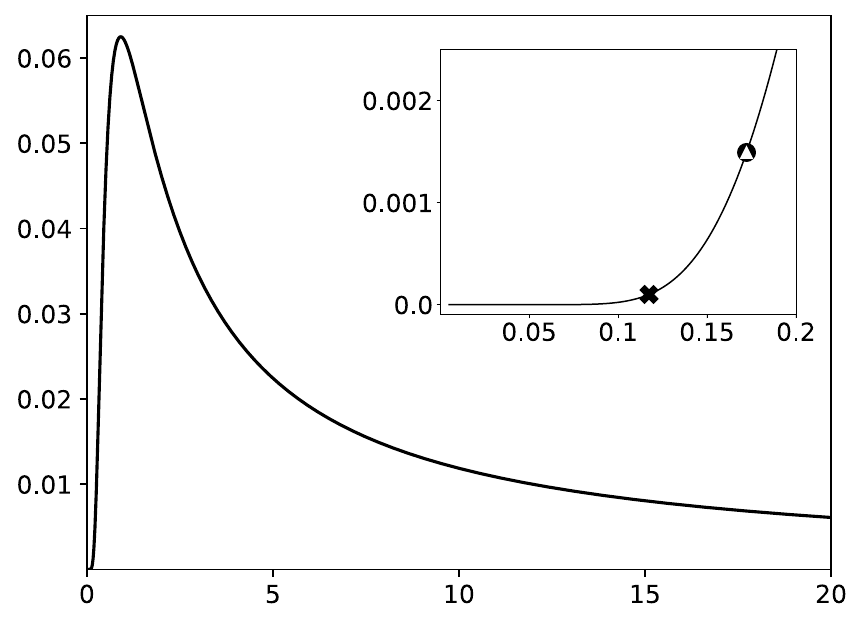}
\put(-3,36){\rotatebox{90}{$\bar{w}$}}
\put(50,-3){$k_BT/\hbar \omega$}
\end{overpic}
\caption{Rescaled minimum work in the thermodynamic limit versus initial temperature $T$. The inset shows an enlargement of the exponential vanishing of work at low initial temperatures. Symbols denote state-of-the-art values for qubits on various contemporary quantum computers, including superconducting qubits (black circle), neutral atom qubits (white triangle), and the trapped ion qubits (black cross).}
\end{figure}

Note that $\overline{W}$ goes to zero as it should for $P_1=1/2$ (at infinite initial temperature, any $U$ will leave $\rho$ unaltered), and for $P_1=0$ (at zero initial temperature, the best you can do is to leave $\rho$ unaltered, i.e., $U=\mathbb{1}$).
It is instructive to rewrite the scaling function $\bar{w}$ in terms of initial temperature $T$
\begin{align}
\label{eq:rescaled_min_W_T}
\bar{w}(T) = \frac{\hbar\omega}{2} \frac{\tanh(\frac{\hbar\omega}{2k_BT})}{e^{\hbar\omega / k_BT} + 1}, 
\end{align}
plotted in Figure \ref{fig:rescaled_W_vs_T}.  In the high-$T$ regime $\bar{w}$ vanishes like $1/T$ while in the low-$T$ regime, $\bar{w}$ vanishes exponentially,  $e^{-\hbar\omega /T}$.  The inset of Figure \ref{fig:rescaled_W_vs_T} shows an enlargement of the low-$T$ behaviour of $\bar{w}$ and marks state-of-the-art values for qubits on various contemporary quantum computers \footnote{Initial temperatures on commonly available quantum devices can be slightly higher than the state-of-the-art values, but nevertheless, still fall within the low-temperature regime.  For example, for currently available IBM quantum processors, $k_BT/\hbar \omega \approx 0.2$, which equates to a rescaled minimum work $\bar{w} = 0.005$ in Figure \ref{fig:rescaled_W_vs_T}.}, including superconducting qubits (black circle) \cite{opremcak2021high}, neutral atom qubits (white triangle) \cite{evered2023high}, and trapped ion qubits (black crosss) \cite{harty2014high}. All of them are in the low $T$ region of the curve, while early NMR qubits are far beyond the full scale of the plot in the high $T$ regime.
 
\section{Implementation}\label{sec:implementation}
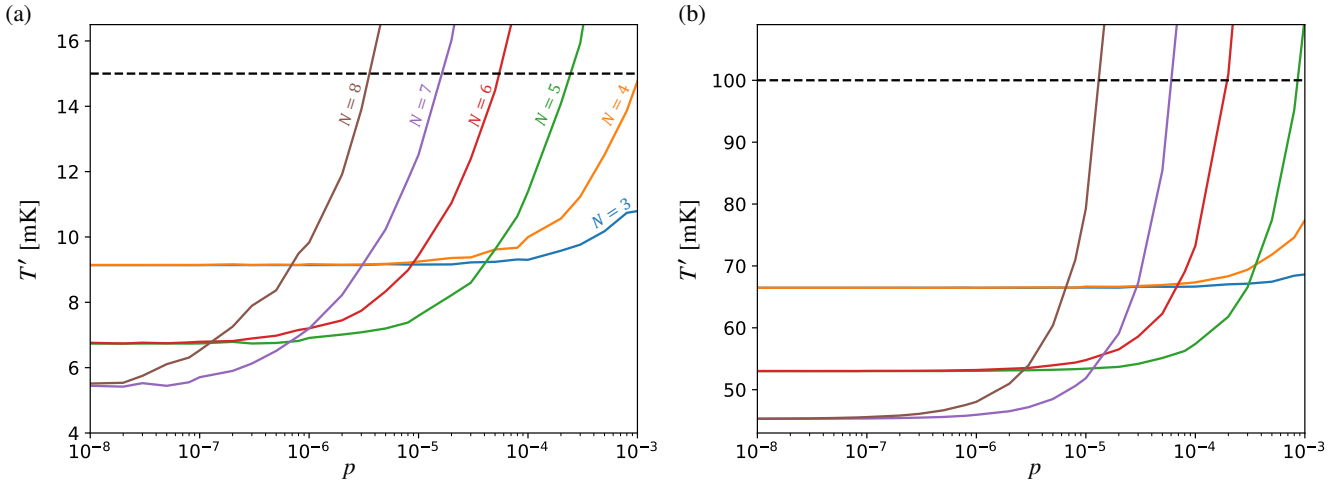
\begin{figure*}[t]
\centering
\begin{overpic}[width=1.99\columnwidth]{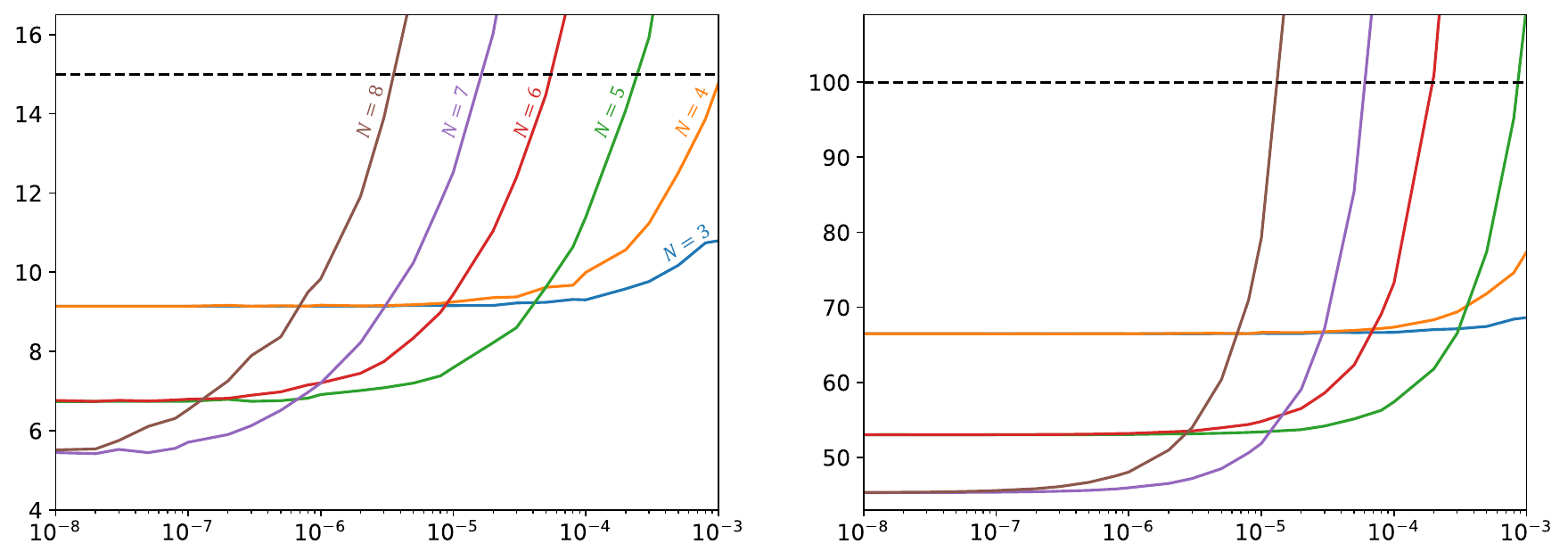}
\put(-3,35){(a)}
\put(49,35){(b)}
\put(-2,15){\rotatebox{90}{$T'$ [mK]}}
\put(23,0){$p$}
\put(49,15){\rotatebox{90}{$T'$ [mK]}}
\put(76,0){$p$}
\end{overpic}
\caption{Final temperatures $T'$ of the target qubit versus noise probability $p$ of the quantum computer for various system sizes $N$ initialized at (a) $T=15$ mK or (b) $T=100$ mK. We assume a value of $\omega = 5$ GHz, which is a typical value for superconducting qubits. The black dashed lines denote initial temperatures of the system. Simulation results from quantum circuits derived from the mirror protocol and executed on a noisy quantum simulator.}
\label{fig:noise_scaling}
\end{figure*}

In order to perform dynamic cooling on quantum computers, the cooling unitary $U$ must be translated into a quantum circuit.  As stated above, there is a large family of unitaries that can achieve maximal cooling, and different quantum circuits will result from different choices of $U$. Near-term quantum computers are noisy, with larger circuits accumulating more errors than smaller ones.  Therefore, from an implementation perspective, cooling unitaries that can be translated into smaller circuits are more desirable (here and below, the size of a quantum circuit refers to the number of constituent elementary one- and two-qubit gates).

Various protocols exist for generating maximally cooling unitaries.  A few protocols of interest include (i) the partner-pairing algorithm (PPA), described in Ref. \cite{schulman2005physical}, (ii) a minimum work protocol, which generates a unitary with minimal work cost, and (iii) a protocol we call the $\emph{mirror}$ protocol.  The mirror protocol is convenient as it can quickly and automatically generate a unique maximal cooling unitary for arbitrarily large $N$ (other protocols can be significantly more computationally difficult, more heuristic, or have degenerate solutions), the downside being that it generates the largest circuits (see Appendix \ref{sec:illustrative_exs} for more details on the various protocols).  The design of a protocol for generating a maximal cooling unitary with minimal circuit size remains an open question for future research.  

It is expected that the size of the cooling circuits will grow exponentially with system size $N$, as the number of states that must be permuted for maximal cooling likewise grows exponentially in $N$ (sub-exponential growth, however, has not been disproved).  See Appendix \ref{sec:qcircuit} for an expanded discussion of the circuit sizes for dynamic cooling.  This scaling has major implications for the practical implementation of dynamic cooling on noisy quantum computers.  Namely, while increasing $N$ increases the theoretically optimal cooling capability, increasing $N$ also increases the depth of the associated circuit, and therefore the accumulated error due to noise. 

Figure \ref{fig:noise_scaling} plots the final temperature of the target qubit versus a noise probability parameter $p$ for various system sizes $N$ for two different initial temperatures.  The results are derived from quantum circuits simulated with a noisy quantum simulator (a classical computer used to simulate the performance of a noisy quantum computer) \footnote{We assume all-to-all connectivity of the qubits.  This means that a two-qubit gate can operate on any pair of qubits in the system.  It should be noted that while trapped ion and neutral atom qubits feature such connectivity, this is generally not the case for superconducting qubits.  Superconducting qubits are usually arranged on a lattice and two-qubit gates can only be applied to neighboring pairs of qubits.  To operate a two-qubit gate on two qubits that are physically separated across the lattice, a series of SWAP gates must be applied, which can dramatically increase circuit depths, further increasing the impact of noise.}, using a noise model based on a depolarizing channel \cite{nielsen2002quantum}, which can be tuned with a single noise parameter $p$ that effectively sets the probability of error.  It is implemented by inserting a random Pauli operator after each gate in the circuit with probability $p$.  The model is commonly used to emulate the performance of circuits on noisy quantum computers as it approximates well the average noise for large circuits \cite{temme2017error, urbanek2021mitigating, dupont2022quantum, oftelie2024simulating}. Furthermore, as it is parameterized by only one parameter, the model facilitates studying the scaling of performance versus noise. The quantum circuits were generated using the mirror protocol, which was chosen because (i) it generates a unique cooling unitary for each system size $N$, providing a fair comparison across various system sizes and (ii) it produces larger circuits than other protocols, meaning that if cooling is possible with the mirror protocol, it will certainly be possible with cooling unitaries better optimized for circuit size. 

We emphasize that these results should only be understood \textit{qualitatively}, since the noise model does not describe the precise noise present on any particular quantum processor.  Moreover, optimizations in terms of selecting a cooling unitary with minimal circuit size and advanced circuit transpilation techniques were not implemented, which would result in shorter, less noisy circuits.   As a result, quantitative conclusions from the plots cannot be drawn; rather, Figure \ref{fig:noise_scaling} serves to reveal trends in how performance scales with noise.

The initial temperature in each plot is indicated with a horizontal black dashed line.  The colored lines indicate the final temperature of the target qubit versus the noise probability $p$ for a range of different system sizes.  Given a system with an odd number of qubits $N$, both plots show that the addition of one more qubit (which theoretically should exhibit identical cooling capability) impairs cooling capability at higher noise probabilities.  While adding more than one qubit to the system increases cooling capability at low noise, we see this can actually decrease cooling capability when noise is sufficiently high.  Furthermore, for a given system size, the noise probability $p$ at which addition of qubits becomes detrimental as opposed to advantageous is smaller when the system is initialized at a lower temperature.  In other words, a system initiated at a lower temperature will be more sensitive to noise.  

We conlcude that in practice, there will be an optimal (finite) number of qubits to use for dynamic cooling, which depends on the level of noise in the quantum hardware as well as the initial temperature of the qubits.  

\section{Demonstration}\label{sec:demonstration}
\begin{figure}\label{fig:cooling_on_brisbane}
\centering
\begin{overpic}[width=0.94\columnwidth]{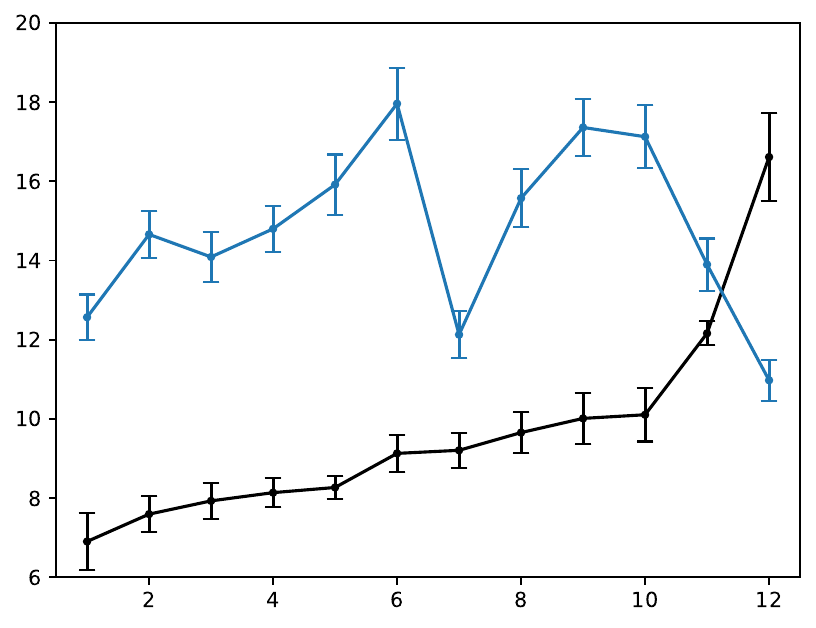}
\put(-3,36){\rotatebox{90}{$T$ [mK]}}
\put(50,-3){Cluster Index}
\end{overpic}
\caption{Initial (black) and final (blue) effective temperatures of various 3-qubit clusters on the IBM quantum computer after dynamic cooling.}
\end{figure}

We demonstrate dynamic cooling with $N=3$ qubits on the IBM quantum computer. Advanced circuit optimization was performed with BQSKit \cite{younis2021berkeley} to reduce the the cooling circuit down to only nine 2-qubit elementary gates.  The circuits were executed on the {\fontfamily{cmtt}\selectfont
ibmq\_brisbane} quantum processor, which contains 127 qubits.   Dynamic cooling was individually performed within twelve different 3-qubit clusters on the chip simultaneously.  Figure \ref{fig:cooling_on_brisbane} plots the presumed initial temperature of each target qubit (black) and the final temperature of each target qubit (blue) after dynamic cooling was executed within each cluster. The presumed initial temperature was computed by executing an empty circuit that only measured the target qubit.  The measurements allowed us to compute the initial population of the excited state of the target qubit, which was then converted into an initial temperature using Eq. \ref{eq:p_to_T}.  This calibration circuit was run five separate times, each with 1024 shots, with the black dots denoting the average value and error bars denoting one standard deviation.  The clusters are indexed from 1 to 12, in increasing order of initial temperature.  Dynamic cooling was executed in 36 separate runs, each with 1024 shots, on each of the 3-qubit clusters, with the blue dots denoting the averaged value and error bars denoting one standard deviation.  The final temperature is analogously calculated using Eq. \ref{eq:p_to_T} with measurements of the final population of the excited state of the target qubit after dynamic cooling.  The black and blue curves are drawn to guide the eye in seeing the successful cooling in cluster 12.

The fact that cooling only occurs in the cluster with the highest initial temperature is in line with the trends revealed in Section \ref{sec:implementation}; namely, qubits at lower temperature are more sensitive to noise, making them harder to cool at a given level of noise. These results suggest that dynamic cooling might best be used in a scheme which scans the initial temperatures of the qubits (or some estimate thereof) and only applies dynamic cooling to those qubits above some threshold initial temperature.

As noise levels continue to decrease on quantum computers, larger circuits will become more feasible to execute, allowing dynamic cooling to be scaled up to larger system sizes $N$, thereby cooling the target qubit down to even lower temperatures.  While the rapid growth of circuit size with increasing $N$ poses a challenge, the fact that cooling scales much better with system size in the low-$T$ regime may enable sufficient cooling with moderately low $N$.  Another path forward to ameliorate large circuit size is a suboptimal cooling scheme, whereby giving up a small amount of cooling power results in a large decrement of circuit complexity, as we next describe in Section \ref{sec:suboptimal}.

\section{Suboptimal dynamic cooling at fixed complexity}
\label{sec:suboptimal}
Currently, the biggest hurdle to the success of dynamic cooling on near-term quantum computers is the size of the cooling circuit. While there is a balanced trade-off between the amount of cooling and energy expenditure at low $T$, circuit size appears to grow exponentially with $N$ (see Appendix \ref{sec:qcircuit}).  While this may seem to be an insurmountable obstacle, here we shall see that it can be overcome by relaxing the requirement of optimal (i.e., maximal) cooling and agreeing to achieve a suboptimal final temperature.  In fact, we find that suboptimal cooling can still (ideally) cool the target qubit down to arbitrarily low temperatures by increasing $N$, but at a fixed circuit complexity and with a lower work expenditure as compared to optimal cooling.  The price that needs to be paid is that the scaling of the final temperature with system size will be slower than $1/N$. 

To see this, consider the following suboptimal cooling protocol.  Take a system with a total of $N=n^2$ qubits, divided into $n$ clusters, each containing $n$ qubits.  Cooling can then be executed in two steps, where first, dynamic cooling is performed within each of the $n$ clusters. Assuming we are operating in the low-$T$ regime, this will bring a total of $n$ qubits to $T'\simeq (2/n) T$ (one qubit from each of the $n$ clusters). In the second step, dynamic cooling is performed amongst these $n$ cooled qubits, bringing one of them to $T''\simeq (4/n^2) T=(4/N) T$. While this is less than the maximal cooling $T'\simeq (2/n^2)= (2/N)T$, it only requires cooling unitaries acting in a space of dimension $\sqrt{N}$, drastically reducing the associated circuit complexity.  Here and below, the complexity of the circuit refers to the maximum Hilbert space dimension on which any of the associated cooling unitaries acts.  If we take, for example, $N=9$, optimal cooling requires a circuit to be generated from a unitary acting on a Hilbert space of dimension $2^9$, whereas the suboptimal cooling circuit is generated from unitaries acting only on Hilbert spaces of dimension $2^3$.  This algorithm can be generalised to $N=n^r$, using $r$ steps to obtain a final suboptimal temperature of $T^{(r)}=(2^r/N) T$.  For a fixed dimension of the clusters $n$, and hence for a fixed circuit complexity, this amounts to a cooling that scales as 
\begin{align}
    T^{(r)}=N^{\frac{\ln 2}{\ln n}-1}T.
    \label{eq:r-step-cooling}
\end{align}
For $n>2$ (which is necessary for cooling to occur in the first place) this implies a negative exponent, ensuring that the qubit can be taken to arbitrarily small final temperature, by increasing $N$, affected by increasing the number of cooling steps $r$ \footnote{Similarly in the high $T$ regime one would get $T^{(r)}=N^{\frac{\ln(\pi/2)}{\ln n}-\frac{1}{2}}T $, with the exponent being negative for $n>2$}.  

The key feature here is that cooling of the target qubit can be augmented by increasing $N$ without increasing the complexity of the circuit, which remains fixed.  Note that in suboptimal cooling, the total circuit comprises a number of $n$-qubit sub-circuits equal to $\sum_{k=1}^{r}n^{r-k} \leq r n^{r-1} = r N/n = N \ln N /(n \ln n )$ . Therefore, the while the circuit size still grows with increasing $N$, it does so only quasi-linearly, as opposed to exponentially with $N$, amounting to a major reduction in circuit size for a given total system size $N$. 

\begin{figure}\label{fig:r-step_noise_scaling}
\centering
\begin{overpic}[width=0.94\columnwidth]{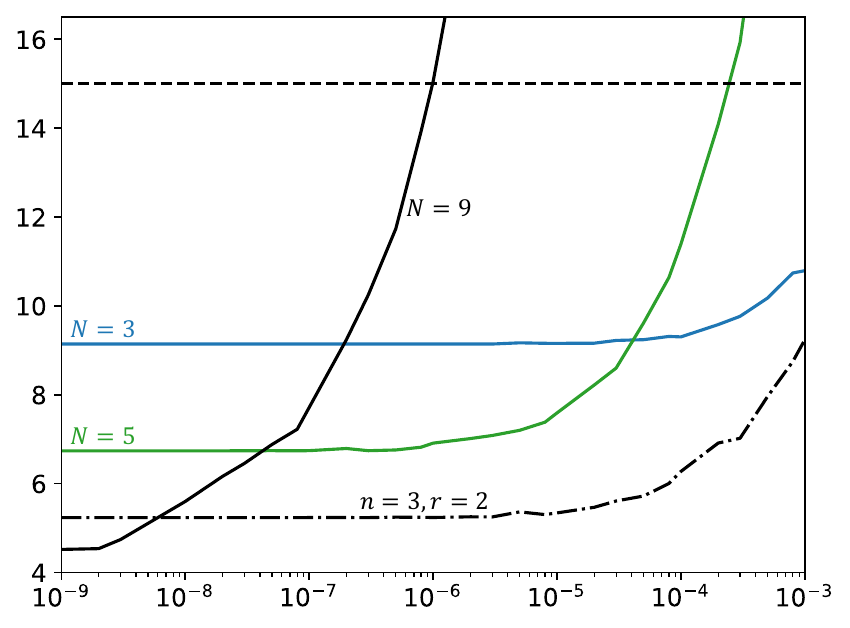}
\put(-3,36){\rotatebox{90}{$T$ [mK]}}
\put(50,-2){$p$}
\end{overpic}
\caption{Comparison of final temperatures for optimal (solid curves) versus suboptimal cooling (dashed-dotted curve) versus noise probability $p$.  We compare optimal cooling with $N=3$ qubits (solid blue), $N=5$ qubits (solid green), and $N=9$ qubits (solid black) to suboptimal cooling with $r=2$ steps of cooling with clusters of size $n=3$, for a total of $N=9$ qubits (dashed-dotted black). An initial temperature of 15 mK is assumed, denoted with the dashed horizontal line.}
\end{figure}

The reduction in circuit complexity of suboptimal cooling significantly increases the feasibility of dynamic cooling on noisy quantum hardware, as evidenced in Figure \ref{fig:r-step_noise_scaling}.  For the same total number of qubits $N=9$, while suboptimal cooling with $n=3$ and $r=2$ (dashed-dotted black curve) relinquishes a small amount of cooling capability at very low levels of noise, it has a significant performance advantage for moderate to high levels of noise compared to optimal cooling (solid black curve).  This is due to the substantially reducted circuit complexity.  Notice how optimal cooling at the same circuit complexity ($N=3$, solid blue curve) achieves significantly less cooling than the suboptimal routine.  Remarkably, the suboptimal cooling routine achieves more cooling than optimal cooling with $N=5$ (solid green curve), even though it has a smaller circuit complexity.  The advantage of the smaller circuit complexity can also be seen by comparing the noise level at which adverse effects begin to impair cooling: reduction in cooling capability begins at a noise level that is an order of magnitude larger for the suboptimal cooling with $n=3, r=2$ versus optimal cooling with $N=5$, confirming that suboptimal cooling is more resilient to noise.

There is also a reduction in the work cost with suboptimal cooling. The total work cost $ W^{(r)}$ is given by the sum of the work costs associated to each step
\begin{align}
    W^{(r)}(P_1,n) = \sum_{k=1}^r n^{r-k} \overline{W}(f_n^{k-1}(P_1),n)
    \label{eq:n-step-work}
\end{align}
where the symbol $f_n^k(x)=(f_n\circ f_n \circ \dots \circ f_n)(x) $ stands for the  $k$-fold application of the function $f_n(x) \doteq P'_1(x,n)$, which we introduce for clarity of notation, and $\overline{W}$ is given in Eq. (\ref{eq:min-W}).  
\begin{figure}\label{fig:r-step_minW}
\centering
\begin{overpic}[width=0.94\columnwidth]{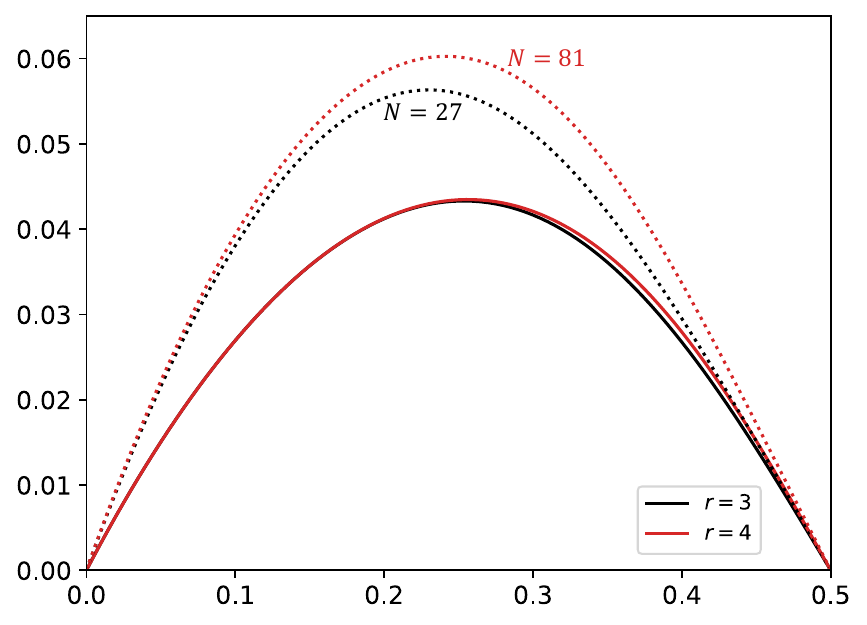}
\put(-3,31){\rotatebox{90}{$W^{(r)}/n^{r}$}}
\put(50,-3){$P_1$}
\end{overpic}
\caption{Rescaled work for $r$-step suboptimal cooling with cluster size $n=3$ (solid curves) and the associated optimal cooling with equal total system sizes (dotted curves) for total system sizes $N=27$ (black) and $N=81$ (red).}
\end{figure}

The solid curves in Figure \ref{fig:r-step_minW} plot the minimal work for suboptimal cooling $W^{(r)}(P_1,n)/n^r$ as a function of $P_1$ for $n=3$, with either $r=3$ (black) or $r=4$ (red). The according values for work at optimal cooling $\overline{W}(P_1,n^r)/n^r$ using the same total number of qubits $N=n^r$ are also plotted for reference with dotted curves and corresponding colors. Note how, as anticipated, the work associated to the suboptimal $r$-step cooling is less than the minimal work associated to optimal cooling for the same total system size $N$. (We remark that a different form of suboptimal cooling was also studied in Ref. \cite{allahverdyan2011thermodynamic}, where it was also shown to lead to a dramatically reduced work cost).  Note also that the curves $W^{(r)}(P_1,n)/n^r$ collapse onto a single curve for growing $N$, meaning that for large $N$, $W^{(r)}(P_1,n)$ scales linearly with $N$, namely, $W^{(r)} \simeq N \bar{w}^{(r)}$.
This linear scaling can be understood by noting that the sum in Eq. (\ref{eq:n-step-work}) is dominated by the first term $n^{r-1} \overline{W}(P_1,n)$, which is upper-bounded by $n^r \bar{w}(P_1)$ which is linear in $N=n^r$. Subsequent terms are upper bounded by $n^{r-k} \bar{w}(f_n^{k-1}(P_1))$. While the factor $n^{r-k}=N^{1-k/r}$ is evidently sub-linear, the overall scaling is much slower than that, because the factor $\bar{w}$ is evaluated at points $f_n^{k-1}(P_1)$ that quickly vanish as $k$ grows (note that $\bar{w}(x)\simeq x/2$ for small $x$, Eq. (\ref{eq:rescaled_min_W})).

We have thus shown the non-trivial fact that one can, in principle, cool a target qubit down to arbitrarily low temperature with fixed circuit complexity and at fixed work cost per qubit. The price that must be paid is that of a slower than linear scaling of cooling with system size, Eq. (\ref{eq:r-step-cooling}). A price that is, however, counterbalanced by a smaller energy cost and significantly reduced circuit complexity and therefore, resilience to noise.

\section{Conclusion and Outlook}\label{sec:conc}
In light of the major developments in quantum technology, which have moved contemporary quantum computers into the low-$T$ regime, we have re-examined dynamic cooling as an effective technique for cooling qubits beyond what is practically achievable with direct, physical cooling methods.  We found an analytic expression for the minimum final temperature $T'$ that can be achieved for the target qubit as a function of the intial temperature $T$. We explored the high-$T$ and low-$T$ regimes and discovered a crossover from a problematic scaling of $1/\sqrt{N}$ at high $T$ to a much more efficient scaling of $1/N$ at low $T$.  We also proposed an analytic expression for the minimal work cost $\overline{W}$ associated to maximal dynamic cooling, which scales linearly with $N$. In particular, while the work cost vanishes like $1/T$ in the high $T$ regime, it vanishes exponentially in the low $T$ regime as $e^{-\hbar \omega/T}$.  

We then turned to the implementation of dynamic cooling on noisy quantum computers.  We noted that different protocols for generating cooling unitaries will give rise to varying quantum circuits sizes, leaving for future work the problem of finding the cooling unitary with minimal quantum circuit size.  We acknowledge that circuit sizes grow rapidly with system size $N$, and explore the implications with simulations of dynamic cooling on a noisy quantum simulator.  The results indicate that there exists an optimal, finite value of $N$ with which to perform dynamic cooling, dependent on the level of noise on the quantum hardware and the initial temperature of the qubits.  

Despite high levels of noise of current quantum computers, we were nevertheless able to successfully demonstrate dynamic cooling on the IBM quantum processor.  Using a system size of $N=3$, we performed dynamic cooling on twelve separate 3-qubit clusters on the 127-qubit chip, observing cooling in just one of the clusters, which is presumed to have been at a higher initial temperature than all the others.  These results suggest that a prudent approach for implementing dynamic cooling on noisy quantum devices may be to scan the initial temperatures (or an estimate thereof) of all the qubits, and only apply dynamic cooling to those qubits above a threshold temperature.  As noise levels continue to decrease, we expect dynamic cooling will be capable of cooling qubits initialized at lower temperatures, and achieve greater cooling using larger system sizes $N$. 

Because of the superior scaling of dynamic cooling with system size in the low-$T$ regime, it may be sufficient to perform dynamic cooling with few enough qubits to maintain reasonable circuit sizes in near-future quantum devices.  However, to overcome the hurdle posed by the rapid growth of circuit size in the near-term, we proposed an algorithm for suboptimal dynamic cooling, whereby instead of reaching the optimal final temperature for a given $N$, we agree to reach a somewhat higher final temperature at the gain of drastically reduced circuit complexity.  Surprisingly, cooling a target qubit down to arbitrarily low temperatures is still possible, in principle.  While cooling scales more slowly than $1/N$ with this suboptimal routine, the circuit complexity remains fixed with increasing $N$, yielding the ability to increase cooling capability without increasing the complexity cost.  

Recent progress in quantum computing technology has exhibited a slow but steady decrease in noise levels, but a relatively fast increase in the total number of available qubits.  Large numbers of moderately low-noise qubits render the suboptimal cooling described above a very viable scheme for the near-future quantum devices.  Furthermore, given the demonstration of cooling with a 3-qubit cluster on a considerably noisy quantum processor, there is hope that suboptimal cooling with cluster sizes of $n=3$ could soon be a realizable path for cooling.  It should be noted, however, that such schemes may require the connectivity of qubits to be re-considered in superconducting qubit implementations, which usually provide lattice-shaped connectivity.  Instead, a fractal-like network of clustered qubits could greatly facilitate the suboptimal cooling algorithm.  Trapped ion and neutral atom quantum computers, which provide all-to-all qubit connectivity, are better suited for both optimal and suboptimal dynamic cooling.

All our results support the conclusion that in the low-temperature regime, dynamic cooling is much more effective in terms of scaling and energy cost than at high initial temperatures, and is capable of achieving cooling when noise is reduced to low enough levels.  Given that current quantum computers operate in the low $T$ regime (unlike early NMR quantum computers), these results reinstate the interest of dynamic cooling for quantum computing applications.

\textit{Note Added.} After submission of our manuscript, it was brought to our attention that Eqs. \ref{eq:y_Nodd} and \ref{eq:y_Neven} had previously been derived using a majorization technique in the Ph.D. Thesis of Rodr\'{i}guez-Briones \cite{rodriguez2020novel}.

\begin{acknowledgments}
LBO gratefully acknowledges funding from the European Union’s Horizon 2020 research and innovation program under the Marie Skłodowska-Curie grant agreement No 101063316.  This research used resources of the National Energy Research Scientific Computing Center (NERSC), a Department of Energy Office of Science User Facility using NERSC award DDR-ERCAP29585.  This research also used resources of the Oak Ridge Leadership Computing Facility, which is a DOE Office of Science User Facility supported under Contract DE-AC05-00OR22725. We acknowledge the use of IBM Quantum services for this work. 
\end{acknowledgments}

\onecolumngrid
\newpage
\appendix

\section{Illustrative examples of maximal dynamic cooling}
\label{sec:illustrative_exs}
A simple procedure to implement maximal cooling amongst $N$ qubits is as follows: consider listing all $2^N$ states in increasing lexicographic order, i.e., $\ket{0...000}, \ket{0...001}, \ket{0...010}, \ket{0...011}, ..., \ket{1...111} $.  The states in the first half of this list all have the target qubit set to 0 ($i_1 = 0)$, and thus have the lower eigenenergy of $K$, $-\hbar\omega/2$.  To maximize cooling, we therefore need to construct a unitary $U$ that maps the $\frac{2^N}{2}$ states of $\rho$ with the highest occupation probabilities to the states in the first half of this lexicographically ordered list.  We emphasize that, for the case of cooling a single target qubit, the order of probabilities within the first half of the list does not matter; all that matters is that the half of states with the highest probabilities are mapped to the first half of the lexicographically ordered list.  

Amongst the degenerate family of maximally cooling unitaries, there is a special transformation, which can be derived with the so-called partner pairing algorithm (PPA), which was demonstrated to be optimal in the entropy manipulation step of (heat bath) algorithmic cooling protocols. The PPA generates a permutation such that states in increasing lexicographic order have non-increasing probabilities. This automatically ensures that the highest probabilities reside in the first half of the lexicographically ordered list.  Other special transformations we will examine are the \emph{mirror protocol}, which allows for simple preparation of the cooling unitary, and the \emph{minimal work protocol}, which performs maximal cooling with minimal work cost.  These will be explained in more detail through the following illustrative examples with $N=3$ and $N=4$ qubits.

\subsection{Maximal cooling with $N=3$ qubits}
Consider three identical qubits all initialized at the same temperature $T$. Our goal is to maximally cool the first qubit below this temperature with the application of a unitary $U$ on all three qubits.

If one lists the $2^N = 8$ states of the total system in lexicographic order, shown in column 2 of Table \ref{table:3qubits}, one sees that they are listed in order of increasing eigenenergies of $K$: the states in the first half of the list, of the form $\ket{0 \hspace{1 mm} i_2 \hspace{1 mm} i_{3}}$, have eigenenergy $-\hbar\omega/2$, while the states in the second half of the list, of the form $\ket{1 \hspace{1 mm} i_2 \hspace{1 mm} i_{3}}$ have eigenenergy $+\hbar\omega/2$.  Initial occupation probabilities of each state $p_i$ are given in column 3, where $x \equiv P_1$ is the initial occupation probability of the excited state for the target qubit.  Notice that the $\frac{2^N}{2} = 4$ largest occupation probabilities are not in the first half of the list.  Specifically, the 4 highest probabilities are $(1-x)^3, x(1-x)^2, x(1-x)^2, x(1-x^2)$.  However, one of these values appears in the lower half of the list. Using the PPA, maximal cooling can be achieved by reordering the probabilities in non-increasing order, which can be accomplished by swapping the two states $\ket{011}$ and $\ket{110}$.

\begin{table}[h]
  \centering 
\begin{tabular}{| c | c | c | c | c | c | c |}
\hline
$i$ & $\ket{i}$ &  $p_i$ & $\ket{\sigma(i)}$ & $p_{\sigma(i)}$ & $\ket{\sigma_B(i)}$ & $p_{\sigma_B(i)}$ \\
\hline \hline
0 & $\ket{000}$  & $(1-x)^3$  &  \textcolor{gray}{$\ket{000}$} & \textcolor{gray}{$(1-x)^3$} &  {$\ket{001}$} & {$x(1-x)^2$} \\
   \hline
1 & $\ket{001}$  & $x(1-x)^2$  &  \textcolor{gray}{$\ket{001}$} & \textcolor{gray}{$x(1-x)^2$} &  {$\ket{100}$} & \textcolor{gray}{$x(1-x)^2$}\\
   \hline
2 & $\ket{010}$  & $x(1-x)^2$  &  \textcolor{gray}{$\ket{010}$} & \textcolor{gray}{$x(1-x)^2$} &  \textcolor{gray}{$\ket{010}$} & \textcolor{gray}{$x(1-x)^2$}\\
   \hline
3 & $\ket{011}$  & $x^2(1-x)$  &  {$\ket{100}$} & $x(1-x)^2$ &  {$\ket{000}$} & $(1-x)^3$\\
   \hline \hline
4 & $\ket{100}$  & $x(1-x)^2$  &  {$\ket{011}$} & $x^2(1-x)$ &  {$\ket{101}$} & $x^2(1-x)$\\
   \hline
5 & $\ket{101}$  & $x^2(1-x)$  &  \textcolor{gray}{$\ket{101}$} & \textcolor{gray}{$x^2(1-x)$} & {$\ket{111}$} & {$x^3$}\\
   \hline
6 & $\ket{110}$  & $x^2(1-x)$  &  \textcolor{gray}{$\ket{110}$} & \textcolor{gray}{$x^2(1-x)$} &  \textcolor{gray}{$\ket{110}$} & \textcolor{gray}{$x^2(1-x)$}\\
   \hline
7 & $\ket{111}$  & $x^3$  &  \textcolor{gray}{$\ket{111}$} & \textcolor{gray}{$x^3$} &  {$\ket{011}$} & {$x^2(1-x)$} \\
   \hline
\end{tabular}
  \caption{All states of the $N=3$ qubit system listed in lexicographic order (column 2) with their initial occupation probabilities $p_i$ (column 3).  Columns 4 and 6 give various permutations, while columns 5 and 7 give the final occupation probabilities of each state after the respective permutation.  (For better readibility the states that are not being displaced by the permutation are in grey).}\label{table:3qubits}
\end{table}

This permutation can be carried out with a unitary operator $U$, defined in the computational basis, as
\begin{equation}
   U_{011 \leftrightarrow 100} =\begin{pmatrix}
    1&0&0&0&0&0&0&0\\
    0&1&0&0&0&0&0&0\\
    0&0&1&0&0&0&0&0\\
    0&0&0&0&1&0&0&0\\
    0&0&0&1&0&0&0&0\\
    0&0&0&0&0&1&0&0\\
    0&0&0&0&0&0&1&0\\
    0&0&0&0&0&0&0&1\\
    \end{pmatrix}~.
    \label{eq:permU3qubit}
\end{equation}

To gain a better understanding of how this unitary performs optimal cooling, we examine how the population of the excited state of the target qubit changes before and after this transformation.  Recall that $P_1$ and $P'_1$ denote the occupation probability of the target qubit's excited state before and after the application of $U$, respectively. Thus, 
\begin{align}
P_1 &=P_{100}+P_{101}+P_{110}+P_{111}\\
P'_1 &=P'_{100}+P'_{101}+P'_{110}+P'_{111}~,
\end{align}
where $P_{i_1i_2i_3}$ is the occupation probability of the state $\ket{i_1 i_2 i_3}$. Now, $U$ implements a transformation that swaps the state $\ket{100}$ with the state $\ket{011}$. Accordingly, it exchanges the populations of the two states, that is $P'_{100}=P_{011}$, and $P'_{011}=P_{100}$, while leaving all other populations unaltered. It follows that
\begin{equation}
\begin{split}
P'_1 &=P'_{100}+P'_{101}+P'_{110}+P'_{111} \\
&= P_{011}+P_{101}+P_{110}+P_{111} \\
&< P_{100}+P_{101}+P_{110}+P_{111} \\
&= P_1
\end{split}
\label{eq:P'<P}
\end{equation}
because the probability $P_{011}=e^{-\hbar\omega\beta/2}/Z^3$ featuring two excitations is lower than $P_{100}=e^{\hbar \omega\beta/2}/Z^3$ featuring only one excitation. From Eq. (\ref{eq:P'<P}) it follows that 
\begin{equation}
P'_1(P_1,3)= 3P_1^2-2P_1^3~.
\label{eq:y3}
\end{equation}
Thus, for $0<P_1<1/2$ (i.e., $T>0$), we have $P'_1<P_1$, namely the target qubit is cooled. 

Due to the degeneracy in the spectra of $\rho$ and $K$, there exists a family of degenerate unitaries that can achieve maximal cooling.  For example, any unitary of the form in Eq. (\ref{eq:permU3qubit}), but with arbitrary phases replacing the $1$'s, can also achieve maximal cooling.  Furthermore, any unitary that swaps states with equal occupation probabilities (in addition to the swap $\ket{100}\leftrightarrow \ket{011}$) also achieves maximal cooling.  In fact, there even exist unitaries performing maximal cooling that implement permutations with cycle lengths greater than two (nb: a swap is a permutation cycle of legnth two).  An example of such a unitary is given by $\sigma_B$ in column 6 in Table \ref{table:3qubits}, featuring the the single cycle of length $6$ given by $\ket{000}\to\ket{001}\to\ket{100}\to\ket{101}\to\ket{111}\to\ket{011}\to\ket{000}$.  In this case, the probabilities are no longer in non-increasing order.  However, the $4$ largest probabilities reside in the first half of the list, which implies maximal cooling of the target qubit.

In the case of $N=3$ qubits, the mirror protocol and the minimal work protocol use the same cooling unitary as the PPA.  Therefore, we reserve explanation of these two protocols until the next illustrative example with $N=4$ qubits, where all three protocols can generate different maximally cooling permutations.

Finally, we remark that cooling cannot be achieved with a total of $N=2$ identical qubits. If the $2^2 = 4$ states are ordered in increasing lexicographic order, $\ket{00},\ket{01},\ket{10},\ket{11}$, the states are automatically listed in order of increasing eigenenergy of $K$ (i.e., energy of the target qubit), and we see that the two highest probabilities already occupy the first half of the list.  Thus, the target qubit cannot be further cooled.

\subsection{Maximal cooling with $N=4$ qubits}
\label{subsec:4qubits}
We now consider the case of $N=4$ qubits.  As before, we list the $2^N = 16$ states of the total system in increasing lexicographic order, as shown in column 2 of Table \ref{table:4qubits}. Again, this automatically orders the states by increasing eigenenergies of $K$: the states in the first half of the list, of the form $\ket{0 \hspace{1 mm} i_2 \hspace{1 mm} i_{3} \hspace{1 mm} i_{4}}$, have eigenenergy $-\hbar\omega/2$, while the states in the second half of the list, of the form $\ket{1 \hspace{1 mm} i_2 \hspace{1 mm} i_{3} \hspace{1 mm} i_{4}}$, have eigenenergy $+\hbar\omega/2$.  The occupation probabilities of the states $p_i$ are now denoted by symbols in column 4 of Table \ref{table:4qubits} to guide the eye to more quickly recognize patterns, where $\blacksquare \doteq (1-x)^4; \hspace{0.1cm} \blacktriangle \doteq (1-x)^3x; \hspace{0.1cm} \textbf{|} \doteq (1-x)^2x^2; \hspace{0.1cm} \bullet \doteq (1-x)x^3; \hspace{0.1cm} \_ \doteq x^4$.  Again, $x \equiv P_1$ is the initial occupation probability of the excited state for the target qubit. Roughly, the more vertices the symbol has, the higher the probability it represents.  The $\frac{2^N}{2} = 8$ largest occupation probabilities can thus be represented by a set containing one $\blacksquare$, four $\blacktriangle$'s, and three out of the six $\textbf{|}$'s. The energy of the total system $E_i$ for each state is given in column 3, which is relevant for determining permutations that maximally cool a target qubit with minimal work cost.

\begin{table*}
  \centering 
\begin{tabular}{| c | c | c | c | c | c | c | c | c | c |}
\hline
$i$ & $\ket{i}$ & $E_i [\hbar\omega/2]$ & $p_i$ & $\ket{\sigma_{PPA}(i)}$ & $p_{\sigma_{PPA}(i)}$ & $\ket{\sigma_{W}(i)}$ & $p_{\sigma_{W}(i)}$ & $\ket{\sigma_M(i)}$ & $p_{\sigma_M(i)}$ \\
\hline \hline
0 & $\ket{0000}$ & $-4$ & $\blacksquare$ &  \textcolor{gray}{$\ket{0000}$} & \textcolor{gray}{$\blacksquare$} &  \textcolor{gray}{$\ket{0000}$} & \textcolor{gray}{$\blacksquare$}&  \textcolor{gray}{$\ket{0000}$} & \textcolor{gray}{$\blacksquare$} \\
   \hline 
1 & $\ket{0001}$ & $-2$ & $\blacktriangle$ & \textcolor{gray}{$\ket{0001}$} & \textcolor{gray}{$\blacktriangle$} & \textcolor{gray}{$\ket{0001}$} & \textcolor{gray}{$\blacktriangle$}& \textcolor{gray}{$\ket{0001}$} & \textcolor{gray}{$\blacktriangle$} \\
   \hline
2 & $\ket{0010}$ & $-2$ &  $\blacktriangle$ & \textcolor{gray}{$\ket{0010}$} & \textcolor{gray}{$\blacktriangle$} & \textcolor{gray}{$\ket{0010}$} & \textcolor{gray}{$\blacktriangle$}& \textcolor{gray}{$\ket{0010}$}  & \textcolor{gray}{$\blacktriangle$} \\
   \hline
3 & $\ket{0011}$ & $0$ & \textbf{|}  & $\ket{1000}$ & $\blacktriangle$ & \textcolor{gray}{$\ket{0011}$}  & \textcolor{gray}{\textbf{|}} & \textcolor{gray}{$\ket{0011}$}  & \textcolor{gray}{\textbf{|}} \\
   \hline
4 & $\ket{0100}$ & $-2$ &  $\blacktriangle$ & \textcolor{gray}{$\ket{0100}$} & \textcolor{gray}{$\blacktriangle$} & \textcolor{gray}{$\ket{0100}$} & \textcolor{gray}{$\blacktriangle$}& \textcolor{gray}{$\ket{0100}$}   & \textcolor{gray}{$\blacktriangle$} \\
   \hline  
5 & $\ket{0101}$ & $0$ &  \textbf{|} & \textcolor{gray}{$\ket{0101}$} & \textcolor{gray}{\textbf{|}} & \textcolor{gray}{$\ket{0101}$} & \textcolor{gray}{\textbf{|}}& \textcolor{gray}{$\ket{0101}$}  & \textcolor{gray}{\textbf{|}} \\
   \hline  
6 & $\ket{0110}$ & $0$ &  \textbf{|} & \textcolor{gray}{$\ket{0110}$} & \textcolor{gray}{\textbf{|}} & $\ket{1000}$ & $\blacktriangle$ & \textcolor{gray}{$\ket{0110}$}   & \textcolor{gray}{\textbf{|}} \\
   \hline
7 & $\ket{0111}$  & $2$ &  $\bullet$ & $\ket{1100}$ & \textbf{|} & $\ket{1010}$ & \textbf{|} & $\ket{1000}$   & $\blacktriangle$ \\
   \hline \hline
8 & $\ket{1000}$ & $-2$ &  $\blacktriangle$ & $\ket{0011}$ & \textbf{|} & $\ket{0110}$ & \textbf{|}& $\ket{0111}$ & $\bullet$ \\
  \hline
9 & $\ket{1001}$ & $0$ & \textbf{|} & \textcolor{gray}{$\ket{1001}$} & \textcolor{gray}{\textbf{|}} & \textcolor{gray}{$\ket{1001}$} & \textcolor{gray}{\textbf{|}}& \textcolor{gray}{$\ket{1001}$}   & \textcolor{gray}{\textbf{|}} \\
   \hline
10 & $\ket{1010}$ & $0$ &  \textbf{|}  & \textcolor{gray}{$\ket{1010}$} & \textcolor{gray}{\textbf{|}} &{$\ket{0111}$} & $\bullet$ & \textcolor{gray}{$\ket{1010}$}   & \textcolor{gray}{\textbf{|}} \\
   \hline
11 & $\ket{1011}$ & $2$ & $\bullet$  & \textcolor{gray}{$\ket{1011}$} & \textcolor{gray}{$\bullet$} & \textcolor{gray}{$\ket{1011}$} & \textcolor{gray}{$\bullet$}& \textcolor{gray}{$\ket{1011}$}   & \textcolor{gray}{$\bullet$} \\
   \hline 
12 & $\ket{1100}$ & $0$ & \textbf{|} & $\ket{0111}$ & $\bullet$ & \textcolor{gray}{$\ket{1100}$}   & \textcolor{gray}{\textbf{|}} & \textcolor{gray}{$\ket{1100}$}   & \textcolor{gray}{\textbf{|}}\\
   \hline
13 & $\ket{1101}$ & $2$  & $\bullet$ & \textcolor{gray}{$\ket{1101}$} & \textcolor{gray}{$\bullet$} & \textcolor{gray}{$\ket{1101}$} & \textcolor{gray}{$\bullet$} & \textcolor{gray}{$\ket{1101}$}  & \textcolor{gray}{$\bullet$} \\
   \hline
14 & $\ket{1110}$ & $2$ & $\bullet$  & \textcolor{gray}{$\ket{1110}$} & \textcolor{gray}{$\bullet$} & \textcolor{gray}{$\ket{1110}$} & \textcolor{gray}{$\bullet$} & \textcolor{gray}{$\ket{1110}$}  & \textcolor{gray}{$\bullet$} \\
   \hline
15 & $\ket{1111}$ & $4$ &  $\_$  & \textcolor{gray}{$\ket{1111}$} & \textcolor{gray}{$\_$} & \textcolor{gray}{$\ket{1111}$} & \textcolor{gray}{$\_$} & \textcolor{gray}{$\ket{1111}$}   & \textcolor{gray}{$\_$} \\
   \hline
\end{tabular}
  \caption{All states of the $N=4$ qubit system listed in lexicographic order (column 2) with total state energy (column 3) and their initial occupation probabilities (column 4).  Columns 5, 7, and 9 give various permutations, while columns 6, 8, and 10 give the final occupation probabilities of each state after the respective permutation. Occupation probabilities are represented by symbols, where $\blacksquare \doteq (1-x)^4; \hspace{0.1cm} \blacktriangle \doteq (1-x)^3x; \hspace{0.1cm} \textbf{|} \doteq (1-x)^2x^2; \hspace{0.1cm} \bullet \doteq (1-x)x^3; \hspace{0.1cm} \_ \doteq x^4$. (For better readibility the states that are not being displaced by the permutation are in grey).}
  \label{table:4qubits}
\end{table*}

There are a number of permutations that will transform the probabilities in the first half of the list into the 8 largest probabilities, three of which are shown in Table \ref{table:4qubits}.  The first is a permutation generated according to the PPA $\sigma_{PPA}$, which permutes all the probabilities into non-increasing order. Note that there is a degenerate family of permutations that can be generated by the PPA.  One such permutation, given in column 5 of Table \ref{table:4qubits}, features two cycles of length $2$ (i.e. two swaps) given by $\ket{0011}\leftrightarrow\ket{1000}$ and $\ket{0111}\leftrightarrow\ket{1100}$.  

The second permutation, $\sigma_W$, given in column 7 of Table \ref{table:4qubits}, features one of a degenerate family of minimal work protocols, which achieves maximal cooling with minimal work cost.  In short, after maximal cooling is achieved by moving the highest half of probabilities to the top half of the lexicographically ordered list, the minimal work protocol sorts the probabilities within each half of the list separately.  Within each half-list, the highest probability is assigned to the state with the lowest total system energy, the second highest probability is assigned to the state with the second lowest total system energy, etc.  It turns out that in the case of $N=4$ qubits, the PPA also belongs to the family of minimal work protocols, but this is not generally the case.

The third permutation, $\sigma_M$, given in column 9 of Table \ref{table:4qubits}, enacts what we call the $\emph{mirror}$ protocol.  In the mirror protocol, states that have the target qubit set to $0$ and have a total of $k<N/2$ bits set to $0$ are swapped with their mirror image (also, called the negative image).  The idea is that these are the states in the top half of the lexicographically ordered list that have lower probabilities than their mirror-image state in the bottom half of the list.  This is because a state with $k<N/2$ bits set to zero will necessarily have fewer bits set to zero than its mirror image, and thereby have a lower occupation probability.  These mirror-image swaps ensure that all states with the target qubit set to $0$ are assigned a higher probability than their mirror-image state, which necessarily have the target qubit set to $1$.  In turn, this means the highest half of probabilities will reside in the first half of the lexicographically ordered list. The advantage of the mirror protocol is two-fold: (i) the ease with which one can automatically generate the maximally cooling unitary for any system size $N$ and (ii) the protocol generates a single, unique cooling unitary for each system size $N$, as opposed to the PPA and minimum work protocols which can generate a family of degenerate cooling unitaries.  

In the mirror protocol for the case of $N=4$, we seek states that start with $0$ and have $k<N/2 = 2$ total bits set to $0$.  The only state that adheres to these criteria is the state $\ket{0111}$, which we swap with its mirror image: $\ket{1000}$.  Notice that it is not a minimal work protocol as the state in the first half of the list with highest total energy $\ket{0111}$ is not assigned the lowest probability in the top-half of the list.  Note, also, that the permutation $\sigma$ on $N=3$ qubits, given in Table \ref{table:3qubits} is an instance of the mirror protocol, as well as a minimal work protocol.

To convince ourselves that all three permutations in Table \ref{table:4qubits} all perform maximal cooling, we can compute the probability of the excited state of the target qubit $P'_1$ after each transformation.  In this case, $P'_1 = P'_{1000}+P'_{1001}+P'_{1010}+P'_{1011}+P'_{1100}+P'_{1101}+P'_{1110}+P'_{1111}$.  By consulting what these constituent probabilities are after each permutation in Table \ref{table:4qubits}, one finds that in all cases:
\begin{align}
P'_1(P_1,4) =  3P_1^2-2P_1^3~.
\label{eq:y4}
\end{align}
This is exactly the same expression found for the $N=3$ case.  Namely, adding a fourth qubit did not increase the cooling power.  This is a special case of a more general result: There is no cooling gain in going from an odd $N$ to $N+1$.  Adding a fourth qubit, however, has the adverse effect of increasing the complexity of the unitary operation needed to implement the cooling (in general, operators acting on larger Hilbert spaces are more complex).  In fact, for a given system size $N$, different permutations will carry different complexities in terms of their implementation in quantum circuits.  Notice that the permutation $\sigma_M$ in Table \ref{table:4qubits} contains one permutation cycle of length $2$ (i.e., a swap) which acts on all the qubits, while $\sigma_{W}$ contains two swaps, but each swap only acts on three out of the four qubits. Such characteristics of the permutation will alter the complexity of the final quantum circuit, and should therefore be considered from a practical standpoint when implementing dynamic cooling on quantum computers.

Another crucial point is that distinct permutations that achieve maximal cooling are generally accompanied by distinct energy costs. For example, the work accompanying the permutation $\sigma_M$ (see Eq. \ref{eq:work} in the main text) is:
\begin{equation}
\begin{split}
W_{\sigma_M} &= (\hbar\omega/2)[2(\blacktriangle - \bullet) + (-2)(\bullet - \blacktriangle)] \\
&=  2 \hbar\omega [ P_0^3P_1-P_0 P_1^3 ]~.
\end{split}
\end{equation}
Similarly, the work accompanying the permutation $\sigma_{W}$ is 
\begin{equation}
\begin{split}
W_{\sigma_{W}} &= (\hbar\omega/2)[2(\textbf{|} - \bullet) + (-2)(\textbf{|} - \blacktriangle)] \\
&=  \hbar\omega [(P_0)^3P_1-P_0 P_1^3 ]~.
\end{split}
\end{equation}
Note that the minimal work permutation costs half work of the mirror protocol. $\sigma_{W}$ is more energy efficient than $\sigma_M$ while achieving the same cooling power $P'_1(P_1,4)$. Hence $\sigma_{W}$ may be preferable when it comes to practical applications.

\section{Proof that $P'_1(P_1,2s-1) = P'_1(P_1,2s)$}
\label{sec:proof}
Below is the proof that, in general $P'_1(P_1,2s-1) = P'_1(P_1,2s)$ for $s \in \mathbb{N}$. For ease of notation we set $x \equiv P_1$ and $y \equiv P'_1$.  Note that in this notation, $(1-x) \equiv P_0$.
For $N=2s$, we have
\begin{align}
   y(x,N) &= \sum_{0 \leq k \leq \frac{N}{2} - 1} \binom{N}{k}(1-x)^{k}x^{N-k}+ \frac{1}{2}\binom{N}{\frac{N}{2}}(1-x)^{\frac{N}{2}}x^{\frac{N}{2}}\nonumber\\
   &= \binom{N}{0}x^{N}+\sum_{1\leq k \leq\frac{N}{2} - 1} \Bigg[\binom{N-1}{k} + \binom{N-1}{k-1}\Bigg](1-x)^{k}x^{N-k}+  \frac{1}{2}\binom{N}{\frac{N}{2}}(1-x)^{\frac{N}{2}}x^{\frac{N}{2}}\nonumber\\
   &= x\left[\binom{N}{0}x^{N-1}+\sum_{1\leq k \leq \leq\frac{N}{2} - 1} \binom{N-1}{k} (1-x)^{k}x^{N-k-1}\right]+ \sum_{1\leq k \leq \frac{N}{2} - 1}  \binom{N-1}{k-1}(1-x)^{k}x^{N-k} \nonumber + \frac{1}{2}\binom{N}{\frac{N}{2}}(1-x)^{\frac{N}{2}}x^{\frac{N}{2}}\nonumber\\
   &= xy(x,N-1)+ (1-x) \left[\sum_{0\leq q \leq \frac{N}{2} - 1}  \binom{N-1}{q}(1-x)^{q}x^{N-1-q}+  \frac{1}{2}\Big[ \binom{N-1}{\frac{N}{2} - 1} + \binom{N-1}{\frac{N}{2}} \Big] (1-x)^{\frac{N-2}{2}}x^{\frac{N}{2}} \right]\nonumber \\
   &= xy(x,N-1)+ (1-x) \left[\sum_{0\leq q \leq \frac{N}{2} - 1}  \binom{N-1}{q}(1-x)^{q}x^{N-1-q}+  \binom{N-1}{\frac{N}{2} - 1}(1-x)^{\frac{N}{2} - 1}x^{\frac{N}{2}} \right]\nonumber \\
   &= xy(x,N-1)+ (1-x) \left[\sum_{0\leq q \leq \frac{N}{2}}  \binom{N-1}{q}(1-x)^{q}x^{N-1-q} \right]\nonumber \\
   &= xy(x,N-1)+ (1-x) y(x,N-1)  \nonumber \\
   & = y(x,N-1)~.
\end{align}
In the second line we used the identity $\binom{N}{k} = \binom{N-1}{k} + \binom{N-1}{k-1}$.  In the fourth line we used a change of variable $q=k-1$.  In the sixth line we used that fact that $\binom{N-1}{\frac{N}{2} -1} = \binom{N-1}{\frac{N}{2}}$ for even N.

\section{Derivation of $c_{s}\simeq (2/\sqrt{\pi}) \sqrt{s}$}
\label{sec:cs_deriv}

We have:
\begin{align}
    &\sum_{k=0}^{s-1} \binom{2s-1}{k} (2s-1-2k)
    = (2s-1) \sum_{k=0}^{s-1} \binom{2s-1}{k} -2  \sum_{k=0}^{s-1} k \binom{2s-1}{k} 
\end{align}
Using the identity 
\begin{align}
\sum_{k=0}^n \binom{n}{k}=2^n
\end{align}
we obtain:
\begin{align}
    \sum_{k=0}^{s-1} \binom{2s-1}{k}  = \frac{1}{2} \sum_{k=0}^{2s-1} \binom{2s-1}{k} = \frac{1}{2} 2^{2s-1}= 2^{2s-2}
\end{align}
where we used the fact that the binomial coefficient is symmetric with respect to reflection about its point of maximum.

Using the identity 
\begin{align}
    k \binom{n}{k}=n \binom{n-1}{k-1}   
    \label{eq:binomialID2}
\end{align}
we get
\begin{align}
    \sum_{k=0}^{s-1} k \binom{2s-1}{k} &= \sum_{k=1}^{s-1} k \binom{2s-1}{k}
    = (2s-1) \sum_{k=1}^{s-1} \binom{2s-2}{k-1}
\end{align}
Furthermore
\begin{align}
    2^{2s-2}=\sum_{k=0}^{2s-2} \binom{2s-2}{k} = 2 \sum_{k=0}^{s-2}  \binom{2s-2}{k} + \binom{2s-2}{s-1}
\end{align}
where we used the reflection symmetry of the binomial coefficient and kept in mind not to count the mid value $\binom{2s-2}{s-1}$ twice. From the above equation it follows that:
\begin{align}
    \sum_{k=1}^{s-1} \binom{2s-2}{k-1} = \sum_{k=0}^{s-2} \binom{2s-2}{k} =\frac{1}{2} \left(2^{2s-2}- \binom{2s-2}{s-1}\right)  
\end{align}

Summing up:
\begin{align}
    &\sum_{k=0}^{s-1} \binom{2s-1}{k} (2s-1-2k)  \nonumber \\
    &= (2s-1)2^{2s-2} -(2s-1) \left(2^{2s-2}- \binom{2s-2}{s-1}\right) \nonumber \\
    &= (2s-1)\binom{2s-2}{s-1} \nonumber \\
    &= s \binom{2s-1}{s}
\end{align}
where we use Eq. (\ref{eq:binomialID2}) in the last equality. Therefore:
\begin{align}
    c_s &= 2^{2-2s}  \sum_{k=0}^{s-1} \binom{2s-1}{k} (2s-1-2k)  \nonumber \\
    &= 2^{2-2s} s \binom{2s-1}{s} \nonumber \\
    &= 2^{2-2s}\, s\,  a_s\, 
\end{align}
where we used $a_s= \binom{2s-1}{s}$. Using Eq. (\ref{eq:as-large-s}) we get:
\begin{align}
    \ln c_s &\simeq 2 \ln 2 -2s \ln 2 + \ln s - \ln 2 - \ln \sqrt{\pi} -\ln \sqrt{s} +s \ln 4  \nonumber \\
    &= \ln 2 -\ln \sqrt{\pi}+\ln \sqrt{s}
\end{align}
and therefore
\begin{align}
    c_s \simeq  \frac{2}{\sqrt{\pi}}  \sqrt{s}.
\end{align}

\section{Derivation of $ a_s \simeq s\ln 4 + O(\ln s)$}
\label{sec:as_deriv}
Using Stirling's approximation
\begin{align}
    N! \simeq \sqrt{2\pi N} (N/e)^N
\end{align}
one finds 
\begin{align}
    \ln a_s = \ln \binom{2s-1}{s} \simeq & \frac{1}{2}\ln (2\pi) +\frac{1}{2} \ln(2s-1) + (2s-1)\ln(2s-1) -(2s-1) \nonumber \\
    & - \frac{1}{2}\ln (2\pi) - \frac{1}{2} \ln s -  s\ln s +s \nonumber\\
    &  - \frac{1}{2}\ln (2\pi) - \frac{1}{2} \ln (s-1) - (s-)\ln (s-1) + s -1 \nonumber \\
     \simeq &  -\ln (2\pi) - \frac{1}{2} \ln s + s \ln 4  \label{eq:as-large-s} 
 \end{align}
that is $a_s \simeq s \ln 4 + O(\ln s)$ where $O(\ln s)$ stands for terms that scale at most like $\ln s$.

\section{Quantum circuit complexity of dynamic cooling}
\label{sec:qcircuit}
We can compare the circuit sizes between various cooling unitaries by defining a systematic way to construct the quantum circuits from the particular unitary.  To do this, note that every cooling unitary is defined by a set of cyclic permutations between specified states.  The total quantum circuit can be constructed by building a sub-circuit for each permutation cycle and then concatenating all sub-circuits.  

To describe the construction of a sub-circuit for a given permutation cycle, we focus on a swap (a cyclic permutation of length $2$), and later explain how to generalize this procedure to cycles of greater lengths.  Suppose we wish to construct the quantum circuit that swaps the two states defined by bitstrings $b_1$ and $b_2$.  First, we define a Gray code between the two bitstrings, which is an ordered list of bitstrings beginning with $b_1$ and ending with $b_2$ where each intermediate bitstring differs by only one bit from the previous one \cite{nielsen2002quantum}.

We define the length of the Gray code $m$ as the number of bitstrings in the series.  
The length of the Gray code will be equal to one more than the number of qubits that differ between the two bitstrings (also known as the Hamming distance).  Given the Gray code, a quantum circuit that implements the swap can easily be constructed using a series of multi-controlled-X (MCX) gates, controlled on $N-1$ qubits, where $N$ is the number of qubits in the system \footnote{A $k$-controlled-X gate is an $X$ (i.e., NOT) gate controlled on the values of $k$ qubits.}. For a swap with a Gray code of length $m$, the quantum circuit will contain $2m-3$ MCX gates (see Ref. \cite{nielsen2002quantum} and Appendix \ref{sec:qcircuit} for an illustrative example). 

Constructing the sub-circuit for a cyclic permutation with a length greater than $2$ is only slightly more involved.  Consider, for example, the cyclic permutation of length $3$: $\ket{i_1} \rightarrow  \ket{i_2} \rightarrow  \ket{i_3} \rightarrow  \ket{i_1}$.  Let the Gray code length between $\ket{i_1} \rightarrow  \ket{i_2}$ be $m_{1\rightarrow2}$ and the Gray code length between $\ket{i_1} \rightarrow  \ket{i_3}$ be $m_{1\rightarrow3}$.  To generate the sub-circuit for this permutation, we first apply the $2m_{1\rightarrow2}-3$ MCX gates to transform the bitstring $i_1$ to $i_2$. Next, we apply the $2m_{1\rightarrow3}-3$ MCX gates to transform the bitstring $i_1$ to $i_3$. For a general permutation of length $l$, the total number of MCX gates required will be $\sum_i [2m_{1\rightarrow i} -3]$ where $m_{1\rightarrow i}$ is the Gray code length from the first bitstring in the cycle to the $i$th bitstring in the cycle, where $i$ goes from $2$ to $l$.

To construct the entire circuit, it is only necessary to concatenate all sub-circuits for each permutation cycle together in any order.  The total circuit will therefore contain a total number of MCX gates equal to $\sum_c \sum_{i_c}[2m_{i_c} -3]$, where the outer sum runs over all permutation cycles $c$ in the cooling unitary and the inner sum runs over all constituent Gray code lengths $m_{i_c}$ of permutation cycle $c$.

The mirror protocol (see definition in Section \ref{subsec:4qubits}) is, by definition, comprised of swaps between states that differ in every qubit.  Therefore, the Gray code length of each swap for a system of size $N$ will be the maximum length of $m=N+1$.  In general, the minimal work protocols contain permutations between states that do not differ in every qubit. Therefore, we expect circuits generated with the mirror protocol to be more complex than those generated with minimal work protocols.  By tailoring the permutations in the cooling unitary to minimise Gray code lengths, is it possible to minimise circuit sizes.  The trade-off is that designing these sets of permutations is currently a heuristic procedure, whereas the mirror protocol can easily and uniquely generate a maximally cooling permutation for each system size $N$.

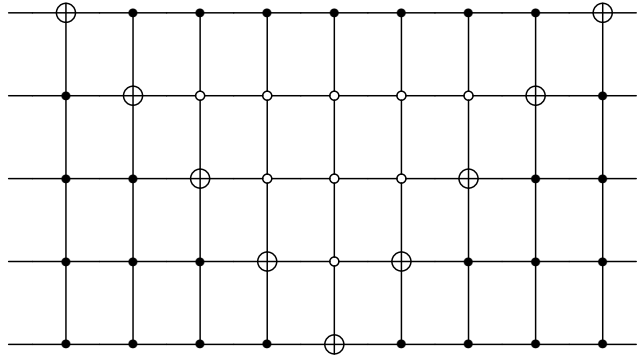
\begin{figure} 
 \centerline{%
    \Qcircuit @C=1em @!R {
        & \qw & \targ & \qw & \ctrl{1} & \qw & \ctrl{1} & \qw & \ctrl{1} & \qw & \ctrl{1} & \qw & \ctrl{1} & \qw & \ctrl{1} & \qw & \ctrl{1} & \qw & \targ & \qw \\
        & \qw & \ctrl{-1} & \qw & \targ & \qw & \ctrlo{1} & \qw & \ctrlo{1} & \qw & \ctrlo{1} & \qw & \ctrlo{1} & \qw & \ctrlo{1} & \qw & \targ & \qw & \ctrl{-1} & \qw \\
        & \qw & \ctrl{-1} & \qw & \ctrl{-1} & \qw & \targ & \qw & \ctrlo{1} & \qw & \ctrlo{1} & \qw & \ctrlo{1} & \qw & \targ & \qw & \ctrl{-1} & \qw & \ctrl{-1} & \qw \\
        &\qw & \ctrl{-1} & \qw & \ctrl{-1} & \qw & \ctrl{-1} & \qw & \targ & \qw & \ctrlo{1} & \qw & \targ & \qw & \ctrl{-1} & \qw & \ctrl{-1} & \qw & \ctrl{-1} & \qw \\
        & \qw & \ctrl{-1} & \qw & \ctrl{-1} & \qw & \ctrl{-1} & \qw & \ctrl{-1} & \qw & \targ & \qw & \ctrl{-1} & \qw & \ctrl{-1} & \qw & \ctrl{-1} & \qw & \ctrl{-1} & \qw   
    }
 }
\caption{Quantum circuit implementing a swap between states $\ket{01111}$ and $\ket{10000}$ using the Gray code shown in Eq. \ref{gray_code}.  Each wire represents a qubit in system.  The circuit is comprised of $2m-3$ MCX gates, where $m=6$ is the length of the Gray code and $N=5$ is the number of qubits in the system.  Open circles with a cross in the MCX gates are the NOT (i.e., Pauli-X) gate, closed circles imply the NOT gate is applied when the corresponding control qubit is in the $\ket{1}$ state, while open circles imply the NOT gate is applied when the corresponding control qubit is in the $\ket{0}$ state.}
\label{fig:gray_code_circuit}
\end{figure}

We demonstrate the generation of the sub-circuit for permutation between the two states $\ket{01111}$ and $\ket{10000}$. To implement the sub-circuit for this swap we let $b_1 = 01111$ and $b_2 = 10000$ and define a Gray code from $b_1$ to $b_2$, such as the following:
\begin{equation} \label{gray_code}
\begin{split}
& 01111 \\
& 11111 \\
& 10111 \\
& 10011 \\
& 10001 \\
& 10000 .
\end{split}
\end{equation}
Here, the length of the Gray code is $m=6$.
To construct the circuit for the swap, we insert one MCX gate to transform each bitstring to the subsequent one in the Gray code.  After insertion of $m-1$ MCX gates, the circuit will successfully transform an input state $b_1$ to $b_2$.  To implement the reverse transformation (since we wish to swap the two states), and uncompute any changes made to other input states not involved in the swap, it is necessary to add the first $m-2$ MCX gates in reverse order.  Thus, a quantum circuit implementing a swap between states with a Gray code of length $m$ will contain $2m-3$ MCX gates.  The quantum circuit implementing the the swap between $b_1$ and $b_2$ using the Gray code in given in Eq. \ref{gray_code} is depicted in Figure \ref{fig:gray_code_circuit}.

Once the quantum circuit has been built with MCX gates, it is necessary to decompose these complex gates into the native gates of the quantum computer (generally these include a two-qubit gate, such as the CNOT gate, and some set of single-qubit gates that render the native gate set universal).  Ref. \cite{barenco1995elementary} describes how such MCX gates can be decomposed into a number of elementary gates (the CNOT gate, and arbitrary 1-qubit gates) that scales quadratically with system size $N$.  However, if we are not concerned about relative phases between the qubits being conserved, the number of elementary gates scales linearly with $N$.  For cooling, we are not concerned about the relative phases between qubits, rather just the populations of each state, and thus, this linearly scaling transformation can be used.

While the number of elementary gates needed for each MCX gate only scales linearly with $N$, unfortunately, the total number of MCX gates in the circuit is expected to grow exponentially with $N$ (since the number of permutations in the cooling unitary is expected to grow exponentially with $N$).  This clearly poses a problem for near-term quantum computers with high levels of noise.  However, since the amount of cooling scales better with system size in the low-$T$ regime, it may be sufficient to cool a target qubit with few enough auxiliary qubits to maintain reasonable circuit sizes on near-future quantum devices with lower levels of noise. It is also possible to suboptimally cool with at a fixed complexity (i.e., elementary gate count) as discussed in Section \ref{sec:suboptimal}.

\begin{figure}\label{fig:cooling_gate_counts}
\centering
\begin{overpic}[width=0.5\columnwidth]{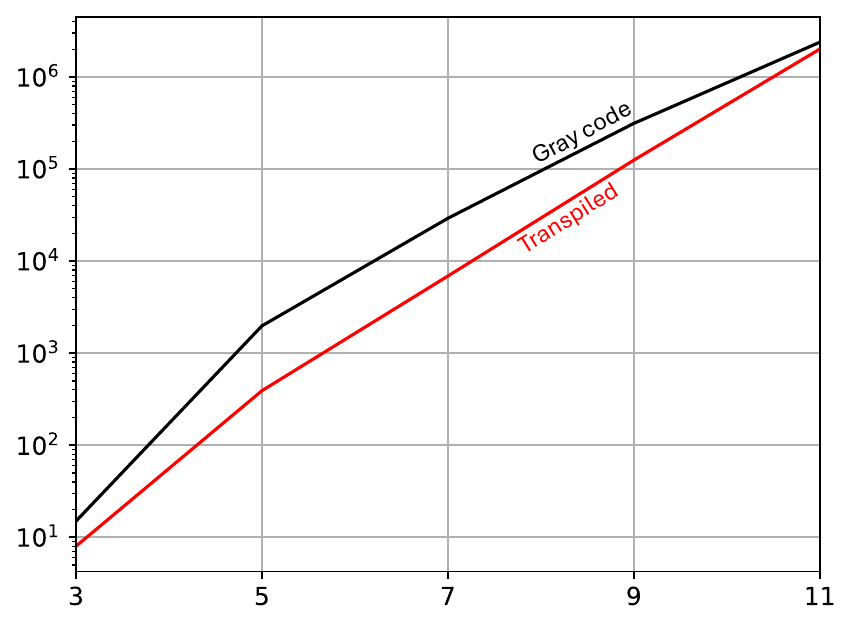}
\put(-3,28){\rotatebox{90}{CNOT count}}
\put(50,-3){$N$}
\end{overpic}
\caption{Number of CNOT gates required for optimal dynamic cooling versus system size N.  The black curve plots the gate count derived from the Gray code method described above, while the red curve plots the gate counts derived from using a circuit transpiler to synthesize the circuit from the input cooling unitary.}
\end{figure}

While it is difficult to derive an analytic expression for the minimal number of elementary gates required for optimal cooling, Figure \ref{fig:cooling_gate_counts} plots a (loose) upper-bound on the number of CNOT gates required for optimal cooling for various system sizes $N$.  The black curve plots the number of CNOT gates calculated using the Gray code method with mirror protocol, as described above.  The red curve plots the number of CNOTs in circuits derived from circuit synthesis using quantum transpilers (such IBM's Qiskit transpiler and the BQSKit transpiler \cite{younis2021berkeley}), which take as input a unitary matrix and output a circuit.  We emphasize that neither of these gate counts describes the minimal gate count for each system size.  There is plenty of room for optimization in terms of selecting a cooling unitary with minimal circuit complexity (indeed, we know the mirror protocol is not optimal for complexity), as well as in terms of circuit transpilation techniques. They do, however, reproduce the expected exponential scaling of CNOT count with system size.  We note that while the circuit transpilation gets extremely computationally expensive as $N$ is increased (we could only go up to $N=11$ in a reasonable amount of compute time), the CNOT count can easily be computed up to any $N$ using the Gray code method.  Therefore, while computing the circuit complexity with the Gray code method will not necessarily give the optimal complexity (as evidenced in Figure \ref{fig:cooling_gate_counts}), it can be useful for quickly comparing complexities between different protocols for large $N$.

\newpage

\twocolumngrid

\bibliography{references}

\end{document}